\documentclass[reprint,aps,pre,showpacs]{revtex4-1}

\usepackage[T1]{fontenc}
\usepackage{bm}
\usepackage{amsfonts}
\usepackage{mathtools}
\usepackage{hyperref}
\usepackage{color}
\usepackage{siunitx}

\newcommand{\V}[1]{{\bm{\mathbf{#1}}}}
\newcommand{\M}[1]{\mathsf{#1}}

\newcommand{\diff}[2]{\frac{\mathrm{d} #1}{\mathrm{d} #2}}

\DeclareMathOperator*{\Var}{Var}
\DeclareMathOperator*{\Cov}{Cov}

\newcommand{\rE}[1]{\left\langle #1\right\rangle_\mathcal{I}}
\newcommand{\sE}[1]{\left\langle #1\right\rangle_\mathcal{E}}

\newcommand{\scov}[2]{\Cov\nolimits_{\mathcal{E}}\left(#1,#2\right)}

\newcommand{\scovnb}[1]{\Cov\nolimits_{\mathcal{E}}#1}
\newcommand{\svarnb}[1]{\Var\nolimits_{\mathcal{E}}#1}
\newcommand{\stcovnb}{\Cov\nolimits^\text{T}_{\mathcal{E}}}

\newcommand{\dd}[1]{\,\mathrm{d} #1}

\newcommand{\bo}[1]{O(#1)}

\newcommand{\Pe}{\mathrm{Pe}}
\newcommand{\Da}{\mathrm{Da}}

\newcommand{\ep}{\varepsilon}
\newcommand{\epm}{{\varepsilon^{-1}}}

\newcommand{\C}[1]{C^{(#1)}}

\DeclarePairedDelimiter\floor{\lfloor}{\rfloor}

\newcommand{\avg}[1]{\left\langle{#1}\right\rangle}
\newcommand{\davg}[1]{\left\langle\left\langle{#1}\right\rangle\right\rangle}

\begin{document}

\title{Stochastic transport in the presence of spatial disorder:\\
fluctuation-induced corrections to homogenization}

\author{Matthew J. Russell}
\email{matthew.russell-3@manchester.ac.uk}

\author{Oliver E. Jensen}
\email{oliver.jensen@manchester.ac.uk}

\affiliation{School of Mathematics, The University of Manchester, Manchester M13
9PL, United Kingdom}

\author{Tobias Galla}
\email{tobias.galla@manchester.ac.uk}
\affiliation{School of Physics and Astronomy, The University of Manchester,
Manchester M13 9PL, United Kingdom}

\date{\today}

\begin{abstract}
Motivated by uncertainty quantification in natural transport systems, we
investigate an individual-based transport process involving particles undergoing
a random walk along a line of point sinks whose strengths are themselves
independent random variables. We assume particles are removed from the system
via first-order kinetics. We analyse the system using a hierarchy of approaches
when the sinks are sparsely distributed, including a stochastic homogenization
approximation that yields explicit predictions for the extrinsic disorder in the
stationary state due to sink strength fluctuations. The extrinsic noise induces
long-range spatial correlations in the particle concentration, unlike
fluctuations due to the intrinsic noise alone. Additionally, the mean
concentration profile, averaged over both intrinsic and extrinsic noise, is
elevated compared with the corresponding profile from a uniform sink
distribution, showing that the classical homogenization approximation can be a
biased estimator of the true mean.

\end{abstract}

\maketitle
\section{Introduction}
Transport processes in natural environments can involve an interplay between
fine-scale disorder in the spatial domain within which transport takes place and
randomness in the transport process itself. Theoretical models that seek to
characterise outcomes in terms of means and covariances must therefore account
for averages over the noise that is intrinsic to the transport process, and
averages over the ensemble of random domains. Spatial averaging (via asymptotic
homogenization or coarse-graining approximations) can be successful in capturing
mean behaviour \cite{davit2013,bruna2015}, but standard techniques often fail to
quantify higher-order uncertainties. Here we use a simple reactive-transport
problem to explore the relationships between intrinsic and spatial averages, and
we present a hybrid homogenization method that predicts mean quantities and
leading-order fluctuations due to the quenched disorder.

While interactions between intrinsic and extrinsic noise appear in applications
ranging from gene expression to epidemic modelling \cite{swain2002, toni2013,
bayati2016, Singh2013}, the problem we address is loosely motivated by
physiology, an area in which predictive models are increasingly taking account
of variability between (and within) individuals in order to inform personalized
medicine \cite{hunter2013}. In the placenta, maternal blood flows in a porous
medium formed by a dense network of branches of villous trees, within which are
capillaries containing fetal blood. Gas and nutrient exchange between mother and
fetus takes place across the syncytiotrophoblast layer coating villous trees.
Oxygen transfer between mother and fetus has previously been approximated using
a simple one-dimensional model in which a chemical species moves via advection
and diffusion past a spatially disordered array of point sinks
\cite{chernyavsky2011transport, chernyavsky2012characterizing}, which take up
the species via zeroth-order kinetics. The concentration of the substance post
disorder average can (in general) be described using a homogenization
approximation; fluctuations around the typical mean behaviour show long-range
spatial correlation and have a structure and magnitude that is sensitive to both
the statistics of the sink distributions and model parameters
\cite{chernyavsky2011transport, chernyavsky2012characterizing}. In some
instances however, the fluctuations can become as great as the mean field itself
and the homogenization approximation fails.

The present problem extends this work in significant respects. First, we treat
the transport as a stochastic process, which enables us to exploit results
derived for zero range processes \cite{levine2005zero,
schadschneider2010stochastic, evans2005nonequilibrium, harris2005current}.
Second, we assume the sinks operate via first-order kinetics and have variable
strength rather than position. These features enable us to derive a hierarchy of
descriptions that exploit the problem's multiscale structure, while remaining
within a linear framework. Third, when the variance in sink strength is
sufficiently small, we show how fluctuations due to the quenched disorder can be
described analytically across a broad range of parameter space of our model
(wider than that accessible to the direct method in
\cite{chernyavsky2011transport,chernyavsky2012characterizing}).
These results can be used to examine systematic differences between averages
over the sink strengths and averages over the intrinsic noise. These
observations also illustrate differences between population-averaged results and
outcomes predicted for an individual, and enable us to quantify the variability
induced by the two distinct sources of disorder in the system.

\section{Model}
\label{sec:model}

We frame our model in a generic manner in order to encompass both
discrete and continuous transport processes. At the discrete level the model
provides a simplified representation of (for example) the Brownian motion of a
virus particle in a mucus film, with diffusive transport interrupted by
adsorption at discrete sites on entangled macromolecules. At the continuum
level, the model describes elements of the transport of a solute in a flow past
an array of sinks, capturing some features of the porous medium encountered by
maternal blood in the placenta, or airflow in a pulmonary acinus. Our main focus
is on determining spatial characteristics of stationary-state particle
distributions.

\subsection{Model definitions and master equation}

\begin{figure*}
    \centering
    \includegraphics[width=0.9\textwidth]{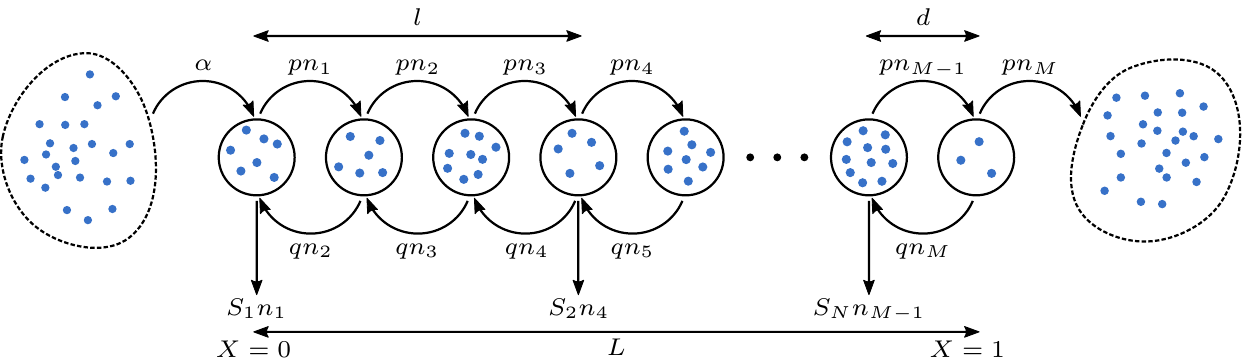}
    \caption{\label{fig:zrp_sinks}Illustration of the stochastic particle
        hopping model: $q n_i$, $p n_i$ are the rates of hopping left, hopping
        right from site $i$; $\alpha$ is the rate of inflow at the left
        boundary; $p n_M$ is the outflow rate at the right boundary; $S_j n_i$
        is the removal rate at sink $j$ (site $i=j\Delta+i_0)$; and $n_i$ is the
        number of particles at the $i$-th site. $\Delta$ is the number of
        regular sites between each pair of sink sites; in this figure $\Delta=3$
        and $i_0=-2$. The long-range dimensionless coordinate $X \in [0,1]$
        spans the physical length $L$ of the domain.}
\end{figure*}

We consider $M$ discrete sites, labelled $i=1,\dots, M$, equally spaced in a
domain of length $L$; see Fig.~\ref{fig:zrp_sinks} for an illustration. The
model describes one species of discrete particles moving in this domain. We
write $n_i(t)$ for the number of particles located at the $i$th site at time
$t$. There is no upper limit on the number of particles that can reside at any
site at any one time. The configuration of the system is determined by the site
occupancies, written as $\V n(t) = (n_1, n_2, \dotsc, n_M)$.

The model operates in continuous time. We assume there is an inflow of particles
at the left boundary with constant rate $\alpha$. Particles do not interact, so
the influx is independent of the occupancy in the first site. In the bulk, each
particle may hop one site to the right or left with rates $p$ and $q$
respectively. The total hopping rate from site $i$ to $i+1$ is then $pn_i$, and
that from $i$ to $i-1$ is $qn_i$. Again there is no interaction between
particles. Particles hopping to the right from the last site leave the system;
the resulting outflow at the end of the chain is $p n_M$.

Particles may also leave the system through a removal process at a subset of $N$
sites that we call sinks; these are located at sites $i_0+\Delta, i_0+2\Delta,
\dots, i_0+N\Delta$, where $N\Delta +i_0 \leq M$. The integer $\Delta$ is the
sink-to-sink distance in units of sites. The particle removal rate at the $j$-th
sink is $S_j n_{j\Delta + i_0}$, $j = 1,\dots, N$, if there are $n_{j\Delta +
i_0}$ particles at the location of the sink. Using $\V e_i$ to denote the unit
$M$-tuple with components $e_{ij} = \delta_{ij}$, the transition rates in the
model are therefore
\begin{multline}
    \label{eqn:transition_rates}
    W_{\V n \to \V m}(\V S) =\\
    \left\{
        \begin{matrix*}[l]
            \alpha & \V m = \V n + \V e_1 &\\
            p n_i & \V m = \V n - \V e_i + \V e_{i+1}, & i=1,\dotsc,M-1\\
            q n_{i+1} & \V m = \V n + \V e_i - \V e_{i+1}, & i=1,\dotsc,M-1\\
            p n_M & \V m = \V n - \V e_M &\\
            S_j n_i & \V m = \V n - \V e_i, & i=j\Delta+i_0,\\
            & &j=1,\dotsc,N\\
            0 & \text{otherwise}. &
        \end{matrix*}
        \right.
\end{multline}
The sink strengths $\V S = (S_1,\dotsc,S_N)$ will be treated as quenched random
variables. They are independently drawn at the beginning, from a distribution
$f(S_i)$ with mean $S_0$, variance $S_0^2\sigma^2$, and then remain fixed during
the transport process.

We denote the conditional probability of finding the system in configuration $\V
n$ at time $t$, given a particular sink strength configuration $\V S$, by $P(\V
n, t | \V S)$. The particles hop according to a continuous-time Markov process
with exponentially distributed waiting times between events. The time-evolution
of the probabilities $P(\V n, t | \V S)$ is governed by the master equation,
\begin{equation}
    \begin{aligned}
        \label{eqn:master_equation_generic}
        \diff{}{t}P(\V n,t | \V S) =
        \sum_\V m \bigl[
        &W_{\V m \to \V n}(\V S)P(\V m,t | \V S)\\
        -& W_{\V n \to \V m}(\V S)P(\V n,t | \V S)
        \bigr],
    \end{aligned}
\end{equation}
with the transition rates as in \eqref{eqn:transition_rates}.
\begin{table}
    \begin{tabular}{|l|c|}
        \hline
        length of domain & $L$ \\
        number of sites & $M$ \\
        number of sinks & $N$ \\
        hopping rates & $p$, $q$ \\
        injection rate & $\alpha$ \\
        mean uptake rate & $S_0$ \\
        \hline
        physical distance between sites & $d=L/(M-1)$ \\
        physical distance between sinks & $\ell=\Delta d$ \\
        number of sites/number of sinks & $\Delta=\frac{M-1}{N+1}$ \\
        advection speed & $u=(p-q)d$ \\
        diffusion coefficient & $D=\tfrac{1}{2}(p+q) d^2$ \\
        \hline
        P\'eclet number & $\Pe=u\ell/D$ \\
        Damk\"ohler number & $\Da = S_0 \ell^2/D$ \\
        inverse number of sinks & $\ep = 1/(N+1)$ \\
        number of sinks/number of sites & $\delta=1/\Delta$ \\
        variance of sink strengths & $\sigma^2$ \\
        concentration scale & $C_0=\alpha Ld/D$ \\
        \hline
    \end{tabular}
    \caption{\label{tab:cases}Summary of model parameters, showing input
        parameters (top), derived quantities (middle) and six independent
        dimensionless parameters (bottom). Continuous descriptions of transport
        and uptake are derived below in a limit in which $\delta \to 0$, $\ep
        \to 0$ and $\sigma\to 0$, with suitable conditions placed on $C_0$,
        $\Pe$ and $\Da$.}
\end{table}
Using the model parameters $p$ and $q$, and the inter-site distance $d=L/(M-1)$
and an inter-sink distance $\ell = d\Delta$, we can identify a mean advection
speed and diffusion
coefficient as
\begin{equation}
    \label{eqn:u_D_p_q}
    u = (p-q) d, \quad D = \tfrac{1}{2}(p+q)d^2.
\end{equation}
For later reference, we introduce a number of dimensionless parameters listed
in Table 1. These include a P\'eclet number, based on the inter-sink distance,
which characterises the relative strength of advection to diffusion,
and a Damk\"ohler number which characterises the relative strength of uptake to
diffusion:
\begin{equation}
    \label{eqn:Pe_Da}
    \Pe = \frac{u\ell}{D} = \frac{2(p-q)\Delta}{p+q}, \quad \Da =
    \frac{S_0\ell^2}{D} = \frac{2S_0 \Delta^2}{p+q}.
\end{equation}
For the mathematical analysis in Sec.~\ref{sec:analytical} below we assume that
the system contains a large number of sites and sinks ($M,N\gg 1$). For later
purposes, it is useful to introduce the inverse number of sinks, $\ep
=1/(N+1)\ll 1$. Our analysis applies for cases in which the sinks are sparsely
distributed relative to the sites; we also introduce the ratio
$\delta=1/\Delta\approx N/M\ll 1$. We will refer to the noise due to the
stochastic hopping as the \emph{intrinsic noise}, and the disorder arising from
the quenched sink strengths as the \emph{extrinsic noise}. We write averages
over the intrinsic noise (i.e., realisations of the stochastic hopping) as
$\rE{\cdots}$ and averages over the extrinsic noise (i.e., the sink strengths)
as $\sE{\cdots}$.

For a fixed realisation $\V S$ of the sink strengths we write
\begin{equation}
    \label{eqn:exp_real}
    \V \rho(t | \V S) \equiv \rE{\V n(t) | \V S} = \sum_\V n \V n(t)
    P(\V n,t | \V S).
\end{equation}
This describes the (intrinsic) mean number of particles at the different sites
at time $t$ for fixed sinks $\V S$. Similarly we introduce an (intrinsic)
covariance between the occupancies $n_i$ and $n_j$, again for fixed sink
strengths $\V S$,
\begin{equation}
    \label{eqn:cov_real}
    \sigma_{ij}(t | \V S) = \rE{n_i(t)n_j(t)|\V S}-\rho_i(t| \V S)\rho_j(t|\V
    S).
\end{equation}
We write $\V \sigma(t | \V S)$ for the resulting covariance matrix.

The mean occupancies post intrinsic average in \eqref{eqn:exp_real} can
further be averaged over the extrinsic uncertainty. We use the following
notation
\begin{equation}
    \label{eqn:exp_sinks}
    \overline{\V \rho}(t) \equiv \sE{\V \rho(t | \V S)} = \int
    \V \rho(t | \V S) F(\V S) \dd \V S,
\end{equation}
writing $F(\V S) \equiv \prod_{i=1}^N f(S_i)$ for simplicity. The shorthand
$\overline{\V \rho}(t) \equiv \sE{\V \rho(t | \V S)}$ is introduced for later
convenience; overbars will be used to indicate averages over the extrinsic
noise. The \emph{total} expectation in (\ref{eqn:exp_sinks}) is an average over
both sources of noise. Analogously, we can introduce
\begin{equation}
    \label{eq:covex}
    \overline{\sigma_{ij}}(t) \equiv \int \sigma_{ij}(t|\V S) F(\V S) \dd \V S,
\end{equation}
and additionally the \emph{extrinsic} covariance,
\begin{multline*}
    \scov{\rho_i(t|\V S)}{\rho_j(t|\V S)} \equiv\\
    \sE{(\rho_i(t|\V S)\rho_j(t|\V S)}-\sE{\rho_i(t|\V S)}\sE{\rho_j(t|\V S)}.
\end{multline*}
The \emph{total} covariance of $n_i(t)$ and $n_j(t)$ is then defined as
\begin{equation}
    \sigma_{ij}^\text{tot}(t) \equiv \davg{n_i(t) n_j(t)}_{\mathcal{I,E}} -
    \davg{n_i}_{\mathcal{I,E}} \davg{n_j}_{\mathcal{I,E}},
\end{equation}
where $\davg{\cdots}_{\mathcal{I,E}}$ stands for the combined average
$\avg{\avg{\cdots}_{\mathcal I}}_{\mathcal E}$. After a modest amount of algebra
one finds
\begin{equation}
    \label{eqn:cov_sinks}
    \sigma_{ij}^\text{tot}(t) = \overline{\sigma_{ij}}(t) + \scov{\rho_i(t|\V
    S)}{\rho_j(t|\V S)},
\end{equation}
an expression of the law of total covariance. The first term in
\eqref{eqn:cov_sinks} is an average of the intrinsic covariance
\eqref{eqn:cov_real} over realisations of the sink strengths. The second term
accounts for correlations between $\rho_i(t|\V S)$ and $\rho_j(t| \V
S)$. These quantities are each obtained from averaging over the intrinsic noise
only, but for a fixed realisation of the sink strengths. They will each depend
on the sink strengths drawn, and can fluctuate together across realisations of
$\V S$.

Finally, we denote quantities in the stationary state of the dynamics by a
superscript `st'. For example, the stationary occupancies, averaged over the
intrinsic noise, will be written as $\rho_i^\text{st}(\V S)$. We will write $\V
\rho^\text{st}(\V S)$ for the vector $(\rho_1^\text{st}(\V S), \dots,
\rho_M^\text{st}(\V S))$.

\section{Numerical simulations}
\label{sec:sims}

In order develop a feeling for the behaviour of the model we first present
numerical simulations. These are carried out in continuous time using the
Gillespie algorithm \citep{gillespie1976general,gillespie1977exact}. We discuss
two sets of simulations. The first set describes a case of densely spaced sinks,
and is for a system of $M=10$ sites with a sink at each site ($\Delta=1$,
$i_0=0$). In the second set, sinks are more sparsely placed, specifically we use
$M=100$ sites, with sinks at every tenth site ($\Delta=10$, $i_0=0$). The
remaining model parameters are $S_0=1$, $p=1$, $q=0.5$ and $\alpha=100$ in both
cases.

\subsection{Densely distributed sinks}

\begin{figure}
    \centering
    \includegraphics[width=\columnwidth]{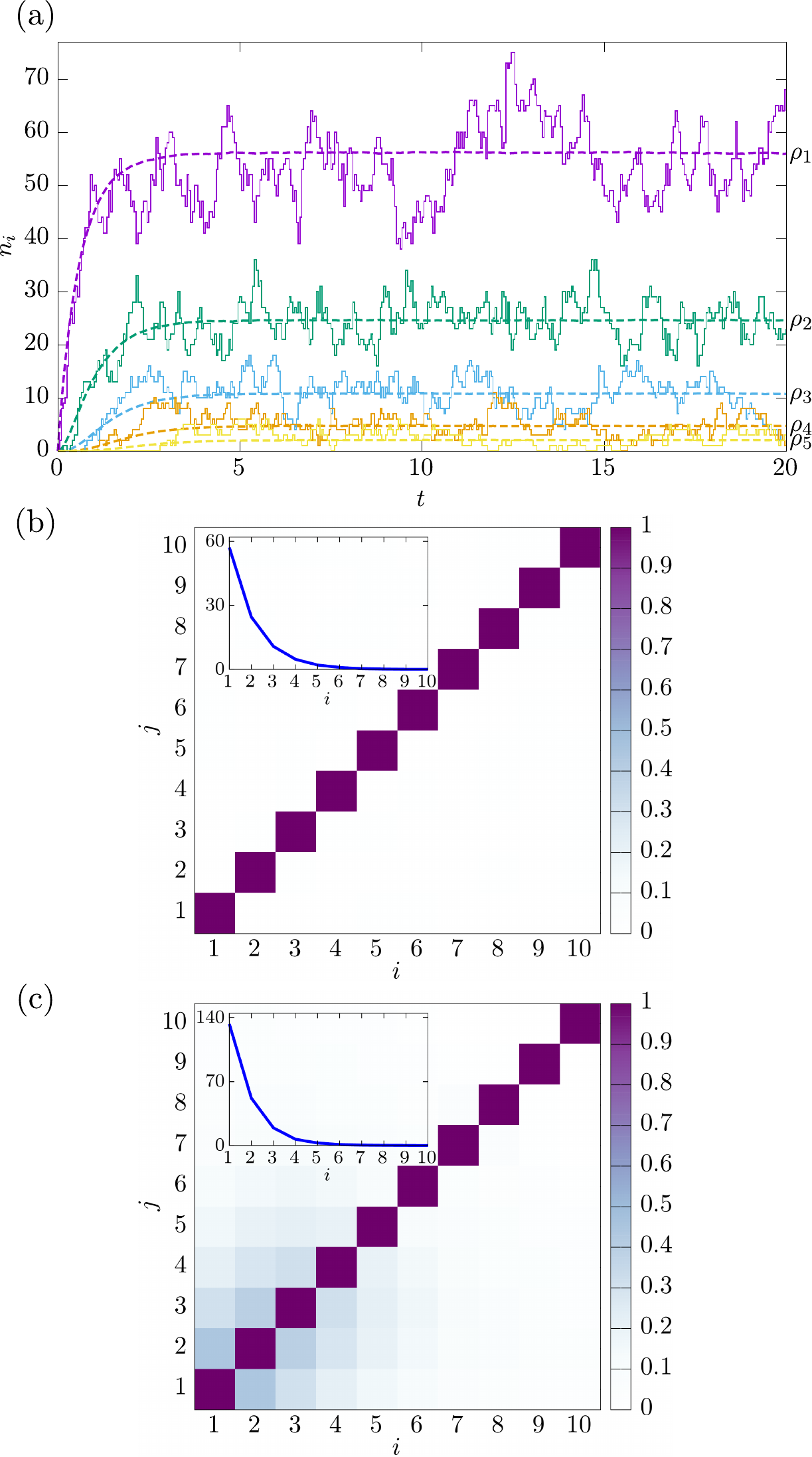}
    \caption{\label{fig:M=10_sims}(a) Dynamics of the system for fixed
        $S_i\equiv 1$. Solid lines show $n_i(t)$, obtained from one single run
        of the Gillespie simulation; dashed lines show $\rho_i(t)$ obtained from
        $10^4$ samples of the intrinsic noise; (b) Normalised stationary
        covariance $\sigma_{ij}^\text{st}/\sqrt{\smash[b]{\sigma_{ii}^\text{st}
        \sigma_{jj}^\text{st}}}$; (c) Total stationary covariance
        $\V{\sigma}^\text{tot,st}$ (normalised as in (b)) for Gaussian sink
        strengths (unit mean, variance $1/16$). Remaining parameters are $M=10$,
        $p=1$, $q=0.5$, $\alpha=100$. Insets show the intrinsic variance
        $\sigma_{ii}$ (panel (b)) and the total variance
        $\sigma_{ii}^{\mathrm{tot}}$ (panel (c)) at each site. In order to
        show their raw magnitude, these are not normalised to unity.}
\end{figure}

We first consider a system with $M=10$ sites, with a sink of strength $S_i = 1$
at each site, resulting in $\Pe=\Da=\tfrac{2}{3}$ for the above choices of $p$
and $q$. There is no extrinsic disorder in this example. Removal is sufficiently
rapid to prevent most particles from reaching ejection at the last site.
Figure~\ref{fig:M=10_sims}(a) illustrates the intrinsic stochasticity of the
dynamics. We show a single realisation of the site occupancies $n_i(t|\V S)$
(solid lines), superimposed onto the mean occupancies $\rho_i(t|\V S)$,
$i=1,\dotsc,5$ , obtained as an average of $10^4$ independent runs. The
intrinsic covariance matrix in the stationary state $\sigma_{ij}^\text{st}$ is
diagonal, see Fig.~\ref{fig:M=10_sims}(b). We show in
Section~\ref{sec:stat_dist} that the occupancies $n^\text{st}_i(\V S)$ and
$n^\text{st}_j(\V S)$ for $i \neq j$ are independent random variables across
realisations of the intrinsic noise whenever $\V S$ is fixed.

In contrast, the total covariance in the stationary state will contain
off-diagonal contributions when there is extrinsic uncertainty, as illustrated
in Fig.~\ref{fig:M=10_sims}(c) for a normal distribution of sink strengths $S_i$
with unit mean and variance $1/16$. We note that a small proportion of the $S_i$
can be expected to be negative in this case; this does not have a significant
bearing on the results in this example. The off-diagonal covariances imply
spatial correlation between the intrinsic means of the occupancies at different
sites across realisations of the quenched disorder.

\subsection{Sparsely distributed sinks}

\begin{figure}[t!]
    \centering
    \includegraphics[width=\columnwidth]{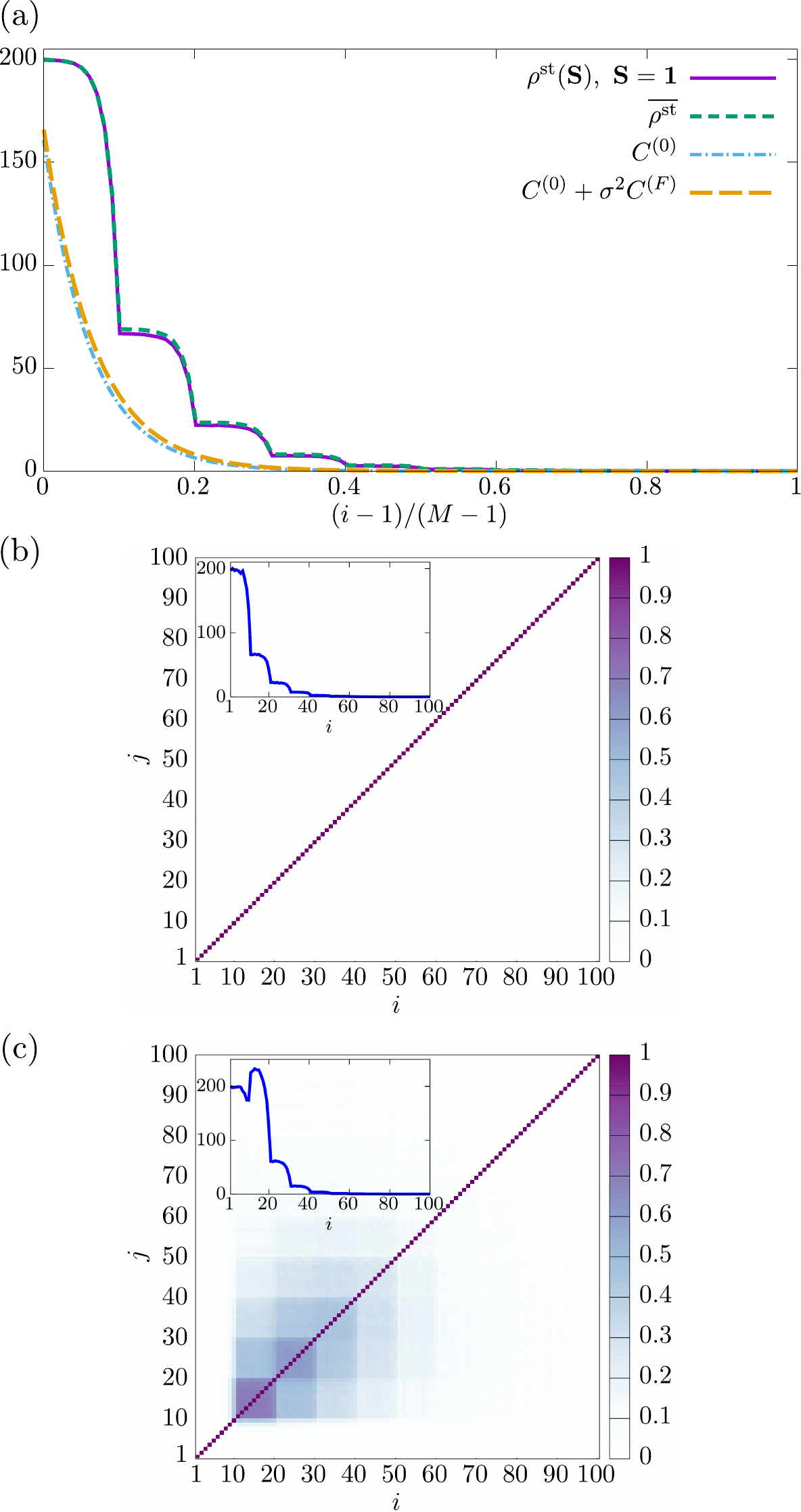}
    \caption{\label{fig:M=100_sims}(a) Stationary mean profiles: intrinsic mean
        with uniform sinks ($\rho^{\text{st}}(\V S)$ with $S_i={1}$, solid) and
        total mean with variable sinks
        ($\overline{\rho^{\text{st}}}$ with $S_i \sim \mathcal{N}(1,1/16)$,
        dashed). Also shown are the leading-order prediction $C^{(0)}$ from
        \eqref{eqn:C0_solution} (dot-dashed), supplemented with its correction
        $\sigma^2 C^{(F)}$ obtained from \eqref{eqn:CF_solution} (large-dashed).
        The normalised site number $(i-1)/(M-1)$ measures distance along the
        entire domain. (b) intrinsic stationary covariance for the case without
        extrinsic noise; (c) total stationary covariance for the case with
        Gaussian disorder. Data are generated from $10^4$ Gillespie runs of the
        stochastic model with $M=100$, $\Delta=10$, $p=1$, $q=0.5$, $\alpha=100$
        and shown for $t=200$. The covariances in (b), (c) are normalised as in
        Figure~\ref{fig:M=10_sims}. Insets show (b) $\sigma_{ii}$ and (c)
    $\sigma_{ii}^{\text{tot}}$.}
\end{figure}

A sparse distribution of sinks introduces a second length scale into the
problem. This can be seen in Fig.~\ref{fig:M=100_sims}(a), which compares the
stationary mean occupancies for fixed sink strengths $S_i\equiv 1$ (i.e., no
extrinsic disorder) and normally distributed sinks ($S_i\sim
\mathcal{N}(1,1/16)$). The parameters we use in this example result in
$\Pe=20/3$ and $\Da=400/3$. The Damk\"ohler and P\'eclet numbers are larger than
in the previous example, i.e.\ sink-to-sink diffusion is weaker than before.
Rapid removal at sinks again prevents most particles from crossing the whole
domain, but the biased hopping is noticeable between each sink, with pronounced
inter-sink staircases superimposed on a decaying profile of particle density.
The total mean occupancy is slightly higher in the case of disordered sinks than
in the case of constant sink strength $S_i\equiv 1$, even though the number of
sinks and their mean strength is the same in both examples; we explore the
origin of this difference below. The intrinsic covariance in the case without
extrinsic disorder ($S_i=1$ for all $i$) is again diagonal, see
Fig.~\ref{fig:M=100_sims}(b), whereas the total covariance with disordered sinks
in Fig.~\ref{fig:M=100_sims}(c) shows long-range spatial correlations and a
multi-scale structure. The intrinsic variance $\sigma_{ii}$ at the different
sites shares the staircase structure of the mean occupancies, see the inset of
Fig.~\ref{fig:M=100_sims}(b). The total variance at the different sites has a
striking non-monotonic form, as shown in the inset of
Fig.~\ref{fig:M=100_sims}(c). This indicates particularly strong variability
immediately downstream of the first sink.

We now explore the origin of the long-range correlations due to fluctuations in
the sink strengths (Figs.~\ref{fig:M=10_sims}(c), \ref{fig:M=100_sims}(c)), the
origins of the elevated total mean occupancy $\overline{\V \rho^\text{st}}$
(Fig.~\ref{fig:M=100_sims}(a)) and seek approximations for the patterns of total
variance. Further simulation data are presented in Figs.~
\ref{fig:homog_vs_monte_carlo}--\ref{fig:lognormal_sinks} below.

\section{Analysis}
\label{sec:analytical}

\begin{figure}
    \centering
    \includegraphics[width=\columnwidth]{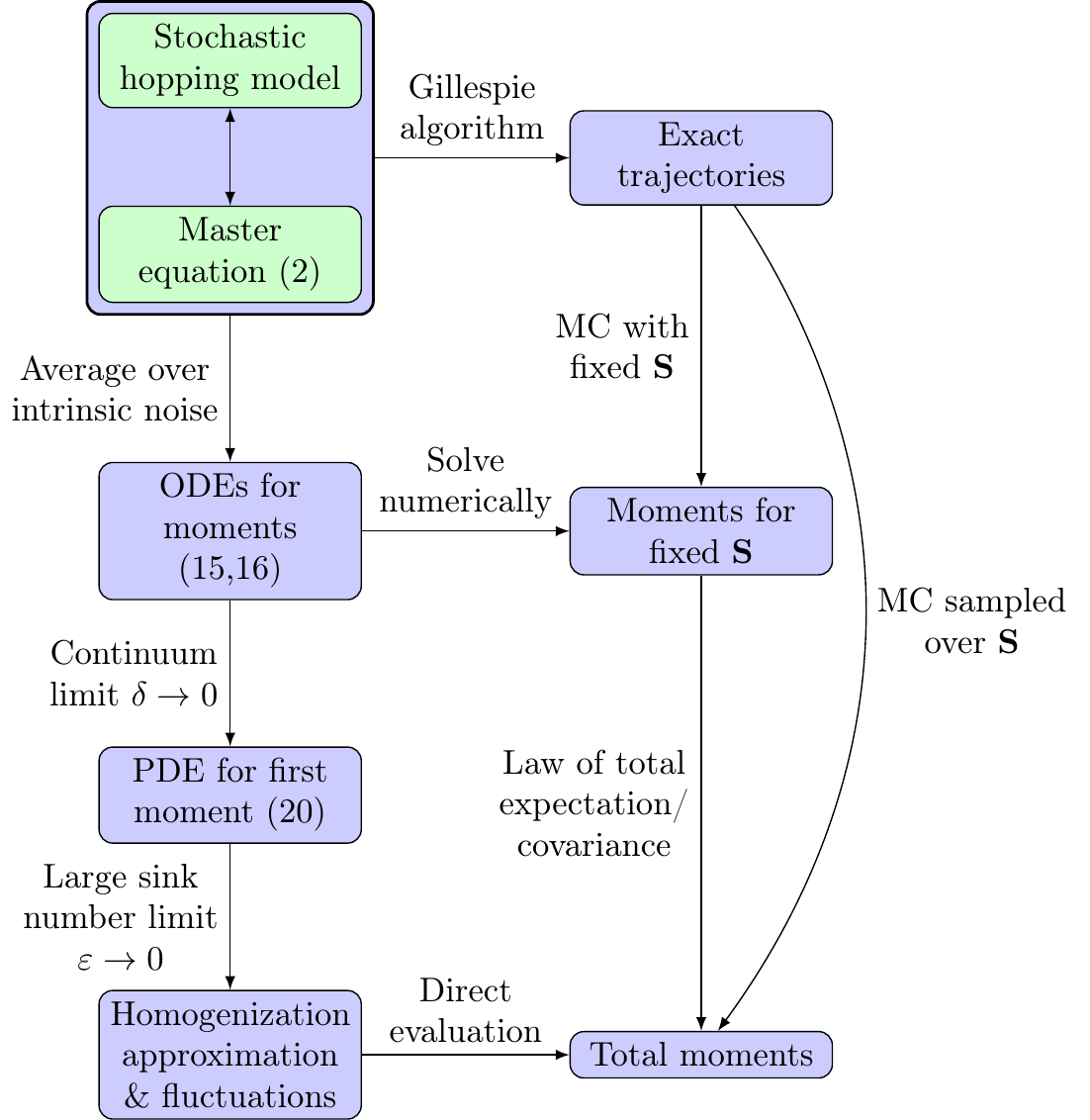}
    \caption{\label{fig:roadmap}Diagram showing the possible pathways between
        various calculation stages in the model. ``MC'' stands for Monte-Carlo.}
\end{figure}

We now proceed with a mathematical analysis of the model. An outline of our
approach is illustrated in Fig.~\ref{fig:roadmap}. We first briefly comment on
the properties of the stationary distribution of the system
(Sec.~\ref{sec:stat_dist}). For a fixed realisation of the sink strengths we
carry out an average over the intrinsic stochasticity and obtain the standard
rate equations for the first and second moments of site occupancies; see
Sec.~\ref{sec:rateeq}. These are ordinary differential equations (ODE), see also
Fig. \ref{fig:roadmap}. Given that there are no interactions between particles
(i.e., reaction rates are linear in the particle numbers), these equations close
and do not involve higher-order moments. In a second step (Sec.~\ref{sec:pde}),
and assuming a sufficiently large injection rate to ensure large particle
occupancy at individual sites and a sparse sink distribution ($\delta \ll 1$),
we take a continuum limit to derive a partial differential equation (PDE) for
the mean occupancy, again for fixed realisations of the sink strengths. The PDE
provides a continuum description of particle transport but retains a discrete
representation of uptake at sinks. Then, assuming a large number of sinks across
the domain ($\ep \ll 1$), we use a stochastic homogenization approach in
Sec.~\ref{sec:hom} to obtain approximations for the total mean and covariance
across the spatial domain. Whereas classical homogenization involves spatial
averaging over a periodic microstructure to derive slow variation over
macroscopic lengthscales, its stochastic analogue goes further by averaging over
a disordered microstructure. In the present case, by assuming the disorder is
weak, we will use the classical formulation as the starting point of a
perturbation expansion in the small sink variance $\sigma^2$. We validate these
theoretical predictions against Monte Carlo simulations in
Sections~\ref{sec:ext}--\ref{sec:mean}. The range of validity of each of these
approximations is assessed as a function of the input parameters of the model in
Sec.~\ref{sec:size}.

\subsection{Stationary distribution, fixed sinks}
\label{sec:stat_dist}

The stochastic model, defined by the transition rates
\eqref{eqn:transition_rates}, is a variant of the open-boundary zero-range
process (ZRP) \citep{levine2005zero, schadschneider2010stochastic,
evans2005nonequilibrium, harris2005current}. It describes non-interacting
particles, and includes particle removal dynamics. The stationary distribution
of the open-boundary ZRP is a product distribution \cite{levine2005zero,
schadschneider2010stochastic}, i.e., in the stationary state the site occupancy
numbers $n^\text{st}_i$, $n^\text{st}_j$ are pairwise independent, and therefore
uncorrelated. This distribution is independent of the initial condition, due to
the ergodicity of the stochastic system. Following Levine \textit{et al.}'s
arguments \cite{levine2005zero, schadschneider2010stochastic}, it can be shown
these properties are left unchanged by the addition of particle removal through
first-order sinks.

Using the results of \cite{levine2005zero}, the stationary distribution of the
model can be written in the form
\begin{equation}
    \label{eqn:joint_factorised}
    P^\text{st}(\V n | \V S) = \prod_{i=1}^M P_i^\text{st}(n_i | \V S),
\end{equation}
where the single-site marginal distributions are Poissonian. Their only
parameters are the stationary mean occupancies $\rho_i^\text{st}(\V S) =
\rE{n_i^\text{st} | \V S }$, for $i = 1, \dotsc, M$. We have
\begin{equation}
    \label{eqn:single_site_marginal}
    P_i^\text{st}(n_i | \V S) =
    \frac{\left[\rho_i^\text{st}(\V S)\right]^{n_i}}{n_i!}
    \exp\left(-\rho_i^\text{st}(\V S)\right).
\end{equation}
Equation~(\ref{eqn:joint_factorised}) can be evaluated if the stationary mean
occupancies $\rho_i^\text{st}(\V S)$ are known. Given the Poissonian nature of
these distributions, we immediately conclude that the (intrinsic) variance at
each site, for a fixed sample of the quenched disorder, equals the mean,
$\sigma_{ii}^\text{st}(\V S) = \rho_i^\text{st}(\V S)$. Furthermore, again for a
fixed sample of the disorder, independence in the stationary state implies that
the second moments factorize, $\rE{n_i^\text{st} n_j^\text{st}} =
\rE{n_i^\text{st}} \rE{n_j^\text{st}}$, as earlier seen for example in
Fig.~\ref{fig:M=100_sims}(b). The total covariance in (\ref{eqn:cov_sinks})
finally becomes

\begin{equation}
    \label{eqn:total_covariance}
    \sigma_{ij}^\text{st,tot} = \overline{\rho_i^\text{st}}\delta_{ij} +
    \scov{\rho_i^\text{st}(\V S)}{\rho_j^\text{st}(\V S)}.
\end{equation}

\subsection{Exact equations for moments, fixed sinks}
\label{sec:rateeq}

The time-evolution of the means and covariances of the site occupancies $n_i$
can be derived directly from the master equation
\eqref{eqn:master_equation_generic}, see for example
\cite{gardiner2009stochastic, van2007stochastic}. It is useful to define the
$M\times M$ matrices $\M A(\V S)$ and $\M B(\V n, \V S)$ as
\begin{subequations}
    \label{eqn:A^1_model}

    \begin{align}
        \begin{aligned}
            A_{ij}(\V S) &\equiv
            \begin{aligned}[t]
                &p(1-\delta_{i,1})\delta_{i,j+1}\\
                &- \left[p(1-\delta_{i,M}) + q(1-\delta_{i,1})\right]
                \delta_{i,j}\\
                &+ q(1-\delta_{i,M})\delta_{i,j-1} - p
                \delta_{i,M}\delta_{i,j}\\
                &+ \delta_{i,j}\textstyle\sum_{k=1}^N \delta_{i,k\Delta + i_0}
                S_k,\\
            \end{aligned}
        \end{aligned}\\
        \begin{aligned}
            B_{ij}(\V n,\V S) &\equiv
            \begin{aligned}[t]
                &p(1-\delta_{i,1})(\delta_{i,j} - \delta_{i,j+1})n_{i-1}\\
                &+ \bigl[p(1-\delta_{i,M})(\delta_{i,j}-\delta_{i+1,j})\\
                &\quad + q(1-\delta_{i,1})(\delta_{i,j}-\delta_{i,j+1}) \bigr]
                n_i\\
                &+ q(1-\delta_{i,M})(\delta_{i,j} - \delta_{i+1,j})n_{i+1}\\
                &+ \delta_{i,1}\delta_{i,j}\alpha + \delta_{i,M}\delta_{i,j} p
                n_M\\
                &+ \delta_{i,j}n_i\textstyle\sum_{k=1}^N \delta_{i,k\Delta +
                i_0} S_k,
            \end{aligned}
        \end{aligned}
    \end{align}
\end{subequations}
for $i,j=1,\dotsc,M$. We note that $\textstyle\sum_{k=1}^N \delta_{i,k\Delta +
i_0} S_k$ is the strength of the sink at site $i$, if there is one; this
expression takes the value zero in absence of a sink at $i$. We also introduce
the vector $\V v$ with entries $v_i = \alpha \delta_{i,1}$. Multiplying the
expressions in \eqref{eqn:master_equation_generic} by $\V n$ and summing over
all configurations $\V n$ yields
\begin{equation}
    \label{eqn:mean_occupancy}
    \diff{}{t} \V \rho(t | \V S) = \M A(\V S) \V \rho(t | \V S) + \V v
\end{equation}
(see Appendix \ref{app:moments} for details). Similarly, for a fixed sample of
the quenched disorder the intrinsic covariances between the occupancies $n_i$
and $n_j$ satisfy \cite{gardiner2009stochastic, van2007stochastic}
\begin{equation}
    \label{eqn:covariance_occupancy}
    \diff{}{t} \V \sigma(t | \V S) = \M A(\V S) \V \sigma(t | \V S) + \V
    \sigma(t | \V S)^T \M A(\V S)^T + \M B(\V \rho, \V S).
\end{equation}

When $\V \rho^\text{st}$ satisfies \eqref{eqn:mean_occupancy} in the
stationary state, it is easily demonstrated that $\sigma_{ij} =
\rho_i^\text{st}\delta_{i,j}$ satisfies \eqref{eqn:covariance_occupancy}.
This is a consequence of Poissonian product form of the stationary distribution
in \eqref{eqn:joint_factorised}.

\subsection{Equations for moments in the continuum limit}
\label{sec:pde}

We now consider the sites arrayed over a continuous spatial domain, and use
\eqref{eqn:mean_occupancy} to derive a PDE for the first moment of the
stochastic transport process at a fixed realisation of sinks. We approximate $\V
\rho_i(t|\V S)$ by a continuous function $C(x,t\vert \V S)$, where $x\in [0,L]$
measures distance along the line of sites. One then has $\rho_i(t | \V S) =
C(x_i,t\vert \V S)$ for $x_i = (i-1)d$, $i=1,\dots,M$. We retain the discrete
locations of the sinks and introduce $S(x) =\sum_{i=1}^N S_i \delta(x-\xi_i)$,
where the $\xi_i=(i_0+i\Delta)d$ are the sink locations in real space.

We first consider the interior of the domain and address the first and last
sites separately. For $i=2,\dots M-1$, \eqref{eqn:mean_occupancy} takes the form
\begin{equation}
    \label{eqn:cont_mean_dim}
    \begin{aligned}
        C_t(x,t) &= p C(x-d, t) - [p+q+S(x)] C(x,t)\\
        &\quad + q C(x+d, t),
    \end{aligned}
\end{equation}
where the subscript $t$ denotes a partial derivative and where we have used the
definition (\ref{eqn:A^1_model}) of the matrix $\M A(\V S)$. We introduce
nondimensional variables, denoted by asterisks, as
\begin{subequations}
    \label{eqn:nondimsclings}
    \begin{gather}
        C^*(x^*,t^*) = C(x,t)/C_0, \quad x^* = x/\ell,\\
        \xi_i^*=\xi_i/\ell, \quad t^*=t/t_0, \quad S_i^*=S_i/S_0,
    \end{gather}
\end{subequations}
where $\ell=d\Delta$ is the physical distance between successive sinks and $t_0
= \ell^2/D$ is the time scale of diffusion between sinks. The quantity $C_0$
drops out in (\ref{eqn:cont_mean_dim}), but it will be defined below. We also
have
\begin{equation}
    \label{eqn:sredef}
    S(x) = S(\ell x^*) = \frac{S_0}{\ell} \sum_{i=1}^N
    S^*_i \delta(x^* - \xi^*_i),
\end{equation}
where the factor $\ell$ is included to ensure that
$\int_{-\infty}^{\infty}\delta(x)\,\dd x = \int_{-\infty}^{\infty}
\delta(x^*)\,\dd x^* = 1$. We substitute \eqref{eqn:nondimsclings} and
\eqref{eqn:sredef} into \eqref{eqn:cont_mean_dim} and expand in $\delta\equiv
1/\Delta=d/\ell \ll 1$. At a fixed number $N$ of sinks, this is valid for large
numbers of sites, $M$. We find
\begin{multline}
    \label{eqn:adre_nondim}
    C^*_{t^*}(x^*,t^*) = -\Pe C^*_{x^*}(x^*,t^*) + C^*_{x^* x^*}(x^* t^*)\\
    - \Da C^*(x^*, t^*)\sum_{i=1}^N S^*_i \delta(x^* - \xi^*_i) + \bo{\delta^3}.
\end{multline}
This advection-diffusion-reaction equation is parameterised by P\'eclet and
Damk\"ohler numbers, defined in \eqref{eqn:Pe_Da}. With multiple
dimensionless parameters in the problem (Table~\ref{tab:cases}), it is important
to distinguish carefully how each behaves when we take the limits of large site
and sink numbers, while preserving low sink density. We analyse this \textit{a
posteriori} in Sec.~\ref{sec:size} below.

The equations at the inflow and outflow boundary sites differ from the bulk and
must be treated separately. Under the scalings \eqref{eqn:nondimsclings} the
inflow boundary equation \eqref{eqn:mean_occupancy} becomes
\begin{align}
    \label{eqn:help}
    \frac{D}{\ell^2} C_0 C^*_{t^*}|_{x^* = 0} = -p C_0 C^*|_{x^*=0} + q C_0
    C^*|_{x^*=\delta} + \alpha.
\end{align}
We can rearrange \eqref{eqn:u_D_p_q} to write the rate constants $p$ and $q$ as
$D/d^2 \pm \tfrac{1}{2}(u/d) = (D/d^2) (1\pm \tfrac{1}{2} \Pe \delta)$
respectively. Expanding \eqref{eqn:help} in powers of $\delta$ and rearranging
gives
\begin{equation}
    \label{eqn:inflow_bc_temp}
     -\Pe C^*|_{x^* = 0} + C^*_{x^*}|_{x^* = 0} + \ep = O(\delta),
\end{equation}
where we have introduced the concentration scale
\begin{equation}
    \label{eq:Cscale}
    C_0 = \frac{\alpha L d}{D} = \frac{2\alpha (M-1)}{p+q}.
\end{equation}
The time derivative is among the $O(\delta)$ terms in (\ref{eqn:inflow_bc_temp})
that are neglected in the limit $\delta \to 0$; this implies that this
approximation may not capture rapid variations in the inlet concentration at
very early times. The leading-order inflow condition is obtained as
\begin{equation}
    \label{eqn:nondim_inflow_bc}
    \Pe C^*|_{x^*=0} - C^*_{x^*}|_{x^* = 0} = \ep.
\end{equation}

Similarly, at the outflow boundary we take the final equation in
\eqref{eqn:mean_occupancy}, write it in terms of the nondimensional continuous
variables, and consider only leading-order terms in $\delta$. We find
\begin{equation}
    \label{eqn:nondim_outflow_bc}
    C^*|_{x^* = \epm} = 0.
\end{equation}
The PDE system (\ref{eqn:adre_nondim},
\ref{eqn:nondim_inflow_bc},~\ref{eqn:nondim_outflow_bc}) provides a convenient
route for approximating conditional means $\V \rho^\text{st}(\V S)$ and, from
\eqref{eqn:total_covariance}, the total covariance. Intersink transport is
governed by the advection-diffusion equation \eqref{eqn:adre_nondim}; the inlet
and outlet conditions are quasi-steady, with advection and diffusion
contributing to the imposed flux $\ep$ in \eqref{eqn:nondim_inflow_bc} and
advection being sufficiently strong to enforce zero concentration at the outlet,
see \eqref{eqn:nondim_outflow_bc}.

Since $\Pe$ and $\Da$ were defined with respect to the intersink distance $\ell$
in (\ref{eqn:Pe_Da}), they appear naturally as parameters in
(\ref{eqn:adre_nondim}). The parameter $\ep$ appears in the domain length
$(0\leq x^*\leq \ep^{-1})$ and the inlet flux. When $\Pe=\Da=0$, the problem has
steady diffusion-dominated solution $C^*=1-\ep x^*$ for which $C^*$ varies by
$O(1)$ across the whole domain, reflecting the balance between inflow and
diffusion across all the sites implicit in (\ref{eq:Cscale}). If we now assume
$\ep \ll 1$ and consider increasing $\Pe$ and $\Da$ from zero, uptake first
becomes important for $\Da=O(\ep^2)$, when $C^*_{x^* x^*}$ balances $\Da C^*$
over a distance $\ep^{-1}$; advection first becomes important for $\Pe=O(\ep)$,
when $C^*_{x^* x^*}$ balances $\Pe C^*_{x^*}$ over a distance $\ep^{-1}$. It
what follows we therefore formally consider the distinguished limit $\ep \to 0$
with $\Pe/\ep$ and $\Da/\ep^2$ remaining $O(1)$ ; these latter quantities are
the P\'eclet and Damk\"ohler numbers defined relative to the domain length $L$.
This ensures that advection, uptake and diffusion are all of comparable
magnitude.

\subsection{Averaging over extrinsic noise}
\label{sec:hom}

We now adopt a homogenization approach, spatially ``smearing'' the discrete sink
locations and averaging over the sink strengths in
(\ref{eqn:adre_nondim},~\ref{eqn:nondim_inflow_bc},~\ref{eqn:nondim_outflow_bc}).
We write the sink strengths as $S_i^*=1+\sigma \hat{S}_i$ where the $\hat{S}_i$
are independent random variables with unit variance. When $\sigma$ is
sufficiently small we may work with $\hat{S}_i\sim \mathcal{N}(0,1)$: a small
number of sink strengths will then be negative, but this is not excluded by our
formalism, and does not change the outcome; alternatively, for larger values of
$\sigma$, we adopt a log-normal distribution.

In the stationary state and dropping asterisks from now on we must solve
\begin{subequations}
    \label{eqn:homog_1st_order_small_x}
    \begin{gather}
        C_{xx} - \Pe C_x = \Da C(x) \sum_{i=1}^N (1+\sigma \hat{S}_i )\delta(x -
        \xi_i)
    \end{gather}
    in 0 $\le x \le \epm$, subject to
    \begin{gather}
        \Pe C|_{x=0} - C_x|_{x=0} = \ep, \quad C|_{x=\epm} = 0.
    \end{gather}
\end{subequations}
Splitting the concentration into its deterministic and fluctuating parts,
$C = \overline C + \sigma \hat C$, where $\overline{C}\equiv \sE{C}$, we can
write
\begin{equation}
    \label{eqn:cbar_chat_eqn}
    \begin{aligned}
        &\overline C_{xx} + \sigma \hat C_{xx} - \Pe \left(\overline C_x +
        \sigma \hat C_x\right) =\\
        & \qquad \Da \sum_{j=1}^N\left(\overline C + \sigma \hat C\right)(1 +
        \sigma \hat S_j) \delta(x-j).
    \end{aligned}
\end{equation}

Averaging \eqref{eqn:cbar_chat_eqn} over the quenched disorder and using the
fact that $\sE{\hat C}=0$ gives
\begin{subequations}
    \label{eqn:cbar_eqn}
    \begin{gather}
        \overline C_{xx} - \Pe \overline C_x = \Da \sum_{j=1}^N
        \left(\overline C + \sigma^2 \sE{\hat C \hat S_j}\right) \delta(x-j),\\
        \Pe \overline C|_{x=0} - \overline C_x|_{x=0} = \ep,\quad
        \overline C|_{x=\epm} = 0,
    \end{gather}
\end{subequations}
while the residual $\hat C$ satisfies
\begin{subequations}
    \label{eqn:chat_eqn}
    \begin{gather}
        \begin{aligned}
            \hat C_{xx} - \Pe \hat C_x = \Da \sum_{j=1}^N \Bigl(& \hat C +
            \overline C \hat S_j \Bigr) \delta(x-j) + O(\sigma),
        \end{aligned}\\
        \Pe \hat C|_{x=0} = \hat C_x|_{x=0},\quad
        \hat C|_{x=\epm} = 0.
    \end{gather}
\end{subequations}
When $\sigma \ll 1$, we may obtain a leading-order approximation to $\overline
C$ by neglecting $\sigma^2 \sE{\hat C \hat S_j}$ in \eqref{eqn:cbar_eqn}, namely
\begin{gather}
    \label{eq:cbarx}
    \overline C_{xx} - \Pe \overline C_x = \Da \sum_{j=1}^N
    \overline C \delta(x-j)
\end{gather}
subject to (\ref{eqn:cbar_eqn}b). We can use this to find $\hat{C}$ in
\eqref{eqn:chat_eqn}, neglecting the $O(\sigma)$ correction in that equation. We
will then return to \eqref{eqn:cbar_eqn} to compute the $O(\sigma^2)$ correction
to $\overline{C}$.

The leading-order approximation for $\overline C$ in (\ref{eq:cbarx}) contains a
periodic array of sinks of fixed strength. At this level we have discarded the
quenched disorder entirely. A classical two-scale asymptotic homogenization
approximation may be adopted for this reduced problem \cite{davit2013}. The
solution is represented as a series $\overline{C}=C^{(0)}(x,X)+\ep
C^{(1)}(x,X)+\ep^2 C^{(2)}(x,X)+\dots$, where we recall that
$\varepsilon=1/(N+1)$ is the inverse number of sinks in the system. The
short-range variable $x$ is treated independently of the long-range variable
$X=\ep x$. We recall that we have dropped asterisks before
\eqref{eqn:homog_1st_order_small_x}, and that $x$ takes values in the interval
$[0,\ep^{-1}]$; the variable $X$ takes values in $[0,1]$. A classical argument,
described for example in \cite{chernyavsky2011transport}, shows that the
leading-order approximation depends only on $X$ and satisfies
\begin{subequations}
    \label{eqn:c0_eqn}
    \begin{gather}
        \ep^2 C^{(0)}_{XX} - \ep\Pe C^{(0)}_X = \Da C^{(0)},
        \quad 0 \le X \le 1,\\
        \Pe C^{(0)}|_{X=0} - \ep C^{(0)}_X |_{x=0} = \ep,\quad
        C^{(0)}|_{X=1} = 0.
    \end{gather}
\end{subequations}
These are derived formally assuming $\Pe=O(\ep)$ and $\Da=O(\ep^2)$, which
ensures a leading-order balance of advection, diffusion and uptake
\cite{chernyavsky2011transport}. This linear problem can be solved directly, and
has solution
\begin{equation}
    \label{eqn:C0_solution}
    \C 0(X) = \frac{\ep e^{\Pe X/2 \ep} \sinh(\epm \phi(1-X))}{\frac{1}{2} \Pe
    \sinh(\epm \phi) + \phi \cosh(\epm \phi)},
\end{equation}
where $\phi \equiv \sqrt{\Da + \Pe^2/4}$. The function $\C0$ varies smoothly
over the length of the domain and provides a leading-order approximation to
$\overline{C}$ in the limit of infinitely many sinks, $\ep\to 0$; higher-order
terms $\C1, \C2, \dotsc$ retain a dependence on $x$ and capture the jump in the
derivative of $\overline{C}$ across each sink.

Comparing (\ref{eq:cbarx}) and (\ref{eqn:c0_eqn}a) illustrates the nature of the
homogenization approach: the discrete sum $\Da \sum_{j=1}^N \overline{C}(x)
\delta(x-j)$ has effectively been replaced by the continuous function $\Da
\overline C(x)$ in order to obtain the leading-order homogenized solution
$C^{(0)}$. This reflects the ``smearing out'' of the sinks, and captures the net
effect of multiple sinks over long length scales. While this ansatz is
appropriate for slowly-varying functions subject to periodic forcing, it cannot
necessarily be adopted more generally.

\begin{figure*}
    \centering
    \includegraphics[width=\linewidth]{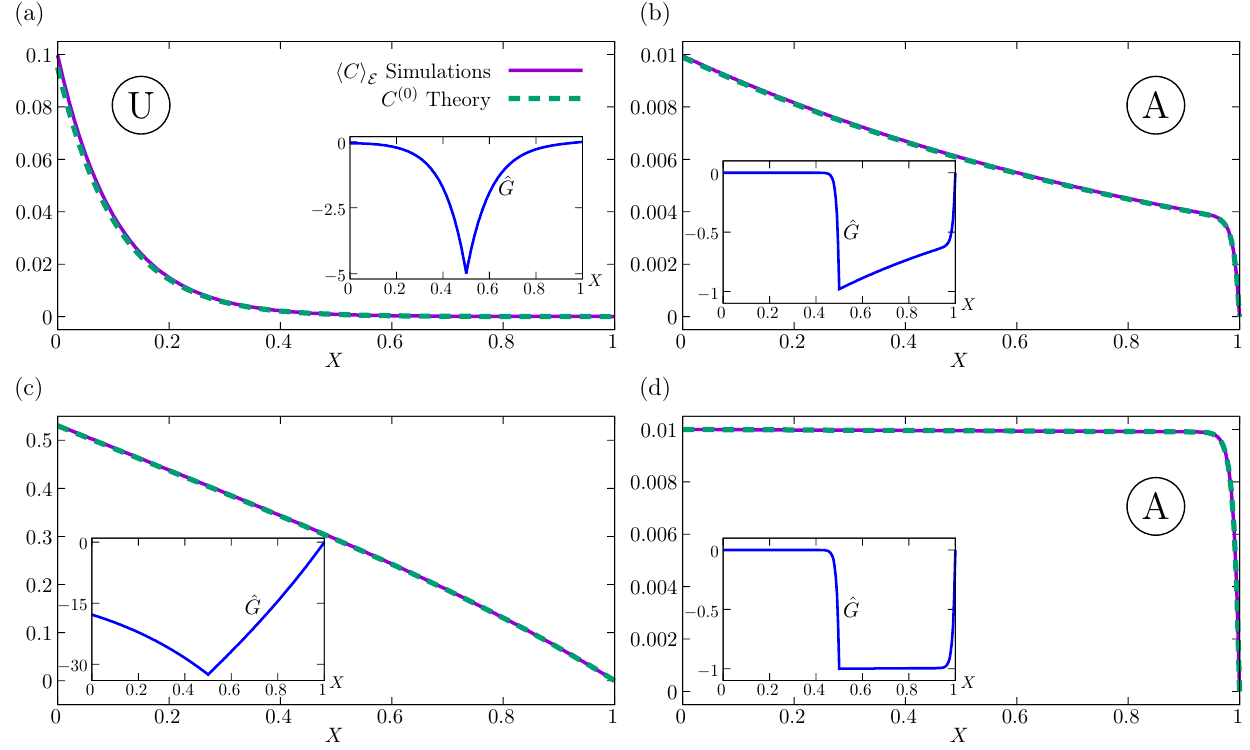}
    \caption{\label{fig:homog_vs_monte_carlo} Comparison between sample means
        $\sE{C}$ calculated from $10^5$ Monte-Carlo simulations of the
        advection-diffusion-reaction equation
        \eqref{eqn:homog_1st_order_small_x} (solid) and the homogenization
        estimate in \eqref{eqn:C0_solution} (dashed). The $99$ sinks are
        normally distributed with $\sigma^2=0.01$. Parameter values $(\Pe,\Da)$
        are (a) $(\ep, \ep)$, (b) $(1,\ep)$, (c) $(\ep,\ep^2)$, (d) $(1,\ep^2)$,
        where the physical interpretation of parameter regime is indicated as a
        circled letter ($U$ stands for an update dominated regime, $A$
        advection dominated cases, and $D$ indicates that diffusion dominates).
        Panel (c) shows a case in which update, advection and diffusion balance
        each other. Insets are the corresponding Green's functions $\hat
        G(X/\ep, 0.5/\ep)$ from (\ref{eqn:greens_func_split},\ref{eqn:G-+}).}
\end{figure*}

Fig.~\ref{fig:homog_vs_monte_carlo} illustrates, for four sets of $(\Pe, \Da)$,
how $\C0$ captures the sample mean over realisations of
\eqref{eqn:homog_1st_order_small_x}. The panels illustrate cases in which (a)
strong uptake leads to rapid decay of the concentration field, (b) elevated
advection displaces the concentration field towards the downstream end of the
domain, (c) advection, diffusion and uptake are in balance across the domain,
and (d) advection is dominant except in a narrow diffusive boundary layer
upstream of the outlet. In panels (a,c), for which $\Pe \ll 1$, diffusion
dominates at the inter-sink scale leading to smooth sample means. In contrast,
when advection becomes significant at the inter-sink scale (as in
Fig.~\ref{fig:M=100_sims}(a), for which $\ep=0.1$), $\C0$ captures the solution
averaged over sinks (with error of $O(\ep)$) but fails to capture its internal
staircase structure. Nevertheless, Figure~\ref{fig:homog_vs_monte_carlo}
illustrates how \eqref{eqn:C0_solution}, derived for $\Pe\sim O(\varepsilon)$
and $\Da\sim O(\varepsilon^2)$, provides a useful approximation across a wide
range of nearby parameter space.

\subsection{Quantifying extrinsic fluctuations}
\label{sec:ext}

We now seek $\hat C$. To solve \eqref{eqn:chat_eqn}, we neglect the $O(\sigma)$
correction that is quadratic in the fluctuations and apply the homogenization
ansatz to the term $\Da \sum_{j=1}^N \hat C \delta (x-j)$, replacing it with
$\Da \hat C(x)$. The perturbations to sink strengths $\hat{S}_j$ vary abruptly
from sink to sink so we retain their discrete form, using $\overline{C}\approx
\C0$ to estimate the strength of each term. This yields the approximate system
\begin{equation}
    \label{eq:duh}
    \begin{aligned}
        \hat C_{xx} - &\Pe \hat C_x - \Da \hat C =
        \Da \sum_{j=1}^N \hat S_j \C0(\ep x) \delta(x-j),
    \end{aligned}
\end{equation}
in $0\le x \le \epm$, subject to (\ref{eqn:chat_eqn}b). It is evident that
$\hat C$ involves multiple independent components, each forced by an individual
sink. This formulation is related to the so-called Duhamel expansion in
stochastic homogenization, for which formal convergence results are available
\cite{bal2011}; similar approaches have been adopted in hydrology
\cite{cushman2002}. The Green's function $\hat{G}(x,y)$ of (\ref{eq:duh},
\ref{eqn:chat_eqn}b) satisfies
\begin{subequations}
    \label{eqn:greens_func_equation}
    \begin{gather}
        \hat G_{xx} - \Pe \hat G_x - \Da \hat G = \delta(x-y), \quad 0 \le x
        \le \epm,\\
        \Pe \hat G|_{x=0} = \hat G_x|_{x=0},\quad
        \hat G|_{x=\epm} = 0,
    \end{gather}
\end{subequations}
and takes the form
\begin{equation}
    \label{eqn:greens_func_split}
    \hat G(x,y) =
    \begin{dcases}
        G_-(x,y) & x \le y,\\
        G_+(x,y) & x > y,
    \end{dcases}
\end{equation}
where
\begin{subequations}
    \label{eqn:G-+}
    \begin{align}
        \label{eqn:G-}
        G_-(x,y) &=
        \frac{ e^{\frac{1}{2} \Pe (x-y)}
        \sinh \left(\phi (y - \epm)\right) g(x) }
        {\phi g(\epm)},\\
        \label{eqn:G+}
        G_+(x,y) &= e^{\Pe(x-y)}G_-(y,x).
    \end{align}
\end{subequations}
We have introduced $g(x)\equiv \Pe\, \sinh(\phi x) + 2 \phi \cosh(\phi x)$.
Like $\C0$, $\hat G$ varies by $O(1)$ with respect to the slow variable $X$,
as illustrated in Fig.~\ref{fig:homog_vs_monte_carlo}(c). The arguments of $\hat
G$, $\Pe\,x$ and $\phi x$, are order unity when $\Pe=O(\ep)$, $\Da=O(\ep^2)$,
and $X=O(1)$. The function $\hat G$ shows more rapid variation with position
when $\Da$ increases, see Fig.~\ref{fig:homog_vs_monte_carlo}(a), or when $\Pe$
increases, see Fig.~\ref{fig:homog_vs_monte_carlo}(b) and (d).

\begin{figure*}
    \includegraphics[width=\linewidth]{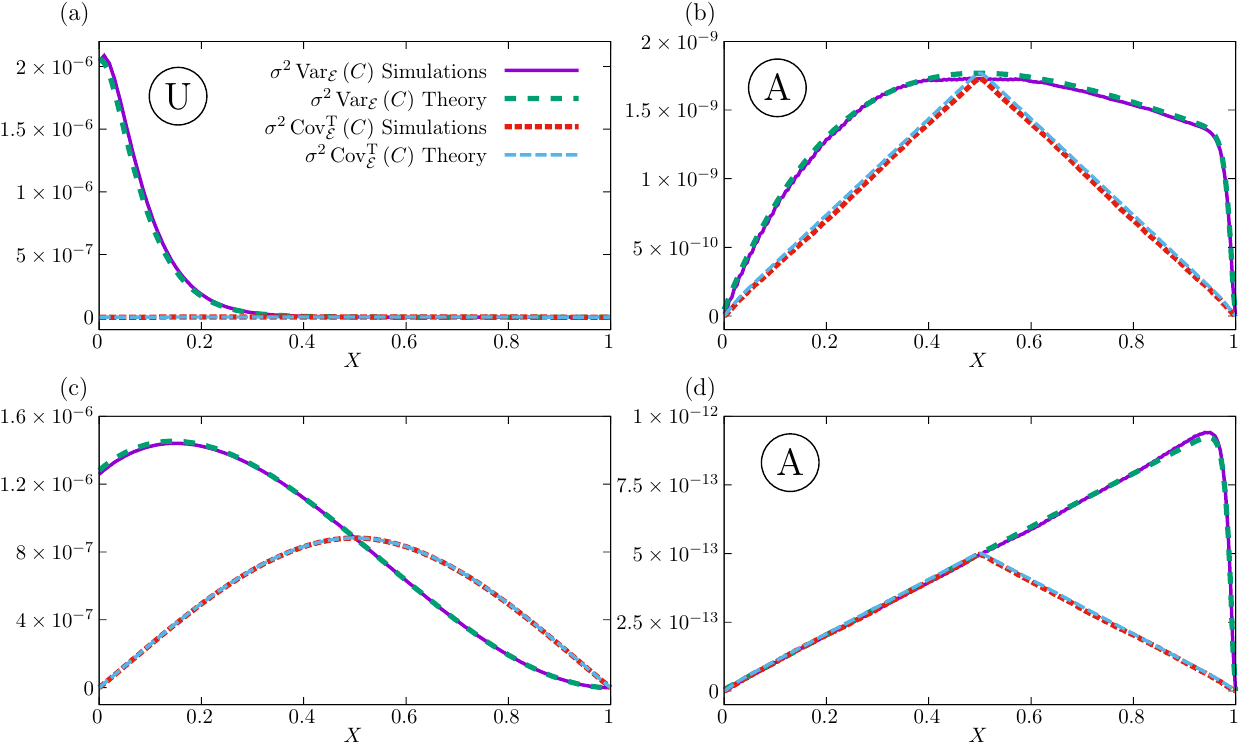}
    \caption{\label{fig:homog_vs_monte_carloB} Comparison between sample
        variance $\sigma^2 \svarnb{(\hat C(X))}$ and transverse covariance
        $\sigma^2 \scovnb{(\hat C(X),\hat C(1-X))}$ calculated from $10^5$
        Monte-Carlo simulations of the ODE \eqref{eqn:homog_1st_order_small_x}
        (solid, thin-dashed) and the theoretical predictions
        (\ref{eqn:C_distribution}, \ref{eqn:cov_theory}) (wide-dashed,
        medium-dashed), using the same parameter values as in
        Figure~\ref{fig:homog_vs_monte_carlo}.}
\end{figure*}

We write $\hat C$ in terms of $\hat G$ and form sums of independent random
variables:
\begin{widetext}
    \begin{equation}
        \label{eqn:hat_C}
        \begin{aligned}
            \hat C(x) &= \Da \int_0^\epm \hat G(x,y) \left[\sum_{j=1}^N \hat S_j
            \C0(\ep j) \delta(y-j)\right] \dd y
            = \Da \sum_{j=1}^i \hat S_j \C0(\ep j) G_+(x,j) + \Da
            \sum_{j=i+1}^N \hat S_j \C0(\ep j) G_-(x,j),
        \end{aligned}
    \end{equation}
    where the integer $i$ is such that $i < x \le i+1$. The resulting sum
    depends on the slow variable $X$ through the slowly varying functions
    $C^{(0)}$ and $G_{\pm}$. Combining the $N$ independent random variables and
    approximating sums with integrals we obtain the approximate distribution of
    $\hat C$, in terms of the long-range coordinate $X$, as
    \begin{equation}
        \label{eqn:C_distribution}
        \hat C(X) \ \ \overset{\mathclap{\text{\tiny approx.}}}{\sim}\ \
        \mathcal{N}\biggl(0, \epm \Da^2 \biggl\{ \int_0^X \C0(X')^2 G_+(\epm X,
        \epm X')^2 \dd X' +\int_X^1 \C0(X')^2 G_-(\epm X, \epm X')^2 \dd X'
        \biggr\} \biggr).
    \end{equation}
\end{widetext}
Using (\ref{eqn:C0_solution}, \ref{eqn:G-+}) and numerically
integrating for different $\Pe$ and $\Da$ yields the variance predictions in
Fig.~\ref{fig:homog_vs_monte_carloB}. These show good agreement with Monte-Carlo
estimates. When advection is strong, the variance increases with distance before
falling to zero at the outlet.

We can also use the approximation for $\hat C$ to compute the transverse
covariances $\stcovnb(\hat C(X)) \equiv \scovnb{(\hat C(X),\hat C(1-X))}$
(derived in Appendix~\ref{sec:corr}). Fig.~\ref{fig:homog_vs_monte_carloB}
confirms that the present analysis captures predictions of Monte Carlo
simulations. Once again the correlation between mean sink occupancies varies
smoothly over the entire length of the domain, despite the fluctuations being
driven over much shorter lengthscales.

\subsection{Influence of fluctuations on mean occupancies}
\label{sec:mean}

We now return to $\overline C$, using \eqref{eqn:hat_C} to evaluate $\sE{\hat C
\hat S_j}$ in \eqref{eqn:cbar_eqn}. Using the fact that $\scovnb{(\hat S_i,\hat
S_j})=\delta_{ij}$, we have
\begin{equation}
\label{eqn:chat_shat}
    \begin{aligned}
        &\sum_{j=1}^N \sE{\hat C \hat S_j} \delta(x-j)\\
        &\enspace = \Da \sum_{j=1}^N \sum_{k=1}^N \sE{\hat S_j \hat S_k} \C0(\ep
            k) \hat G(x,k) \delta(x-j)\\
        &\enspace = \Da \sum_{j=1}^N \sum_{k=1}^N \delta_{jk} \C0(\ep k) \hat
            G(x,k) \delta(x-j)\\
        &\enspace = \Da \sum_{j=1}^N \C0(\ep j) \hat G(x,j) \delta(x-j)\\
        &\enspace \approx \Da \C0(\ep x) \hat G(x,x).
    \end{aligned}
\end{equation}
Because $C_0$ and $\hat{G}$ are smoothly varying functions, it is legitimate to
employ the homogenization ansatz in the final step of (\ref{eqn:chat_shat}).
Thus a refined approximation of $\overline C$ is given by a homogenized version
of (\ref{eqn:cbar_eqn}a) as
\begin{subequations}
    \begin{gather}
        \overline C_{xx} - \Pe \overline C_x - \Da \overline C(x) = \Da^2
        \sigma^2 \C0(\ep x) \hat G(x,x),
    \end{gather}
\end{subequations}
subject to (\ref{eqn:cbar_eqn}b,c). This linear equation can be split into two
parts, $\overline C = \C0 + \sigma^2 \C F$, where $\C0$ satisfies
(\ref{eqn:c0_eqn}), and the correction due to fluctuations in the sinks
satisfies
\begin{subequations}
    \begin{gather}
        \C F_{xx} - \Pe \C F_x - \Da \C F(x) = \Da^2 \C0(\ep x) \hat
        G(x,x),\\
        \Pe \C F|_{x=0} = \C F_x|_{x=0},\quad
        \C F|_{x=\epm} = 0.
    \end{gather}
\end{subequations}
Using $\hat G$ to solve for $\C F$ we obtain,
in long-range coordinates,
\begin{equation}
    \label{eqn:CF_solution}
    \begin{aligned}
        &\C F(X) = \Da^2 \epm \times\\
        & \ \int_0^1 \C0(Y) \hat G(\epm X, \epm Y) \hat G(\epm Y, \epm Y) \dd Y.
    \end{aligned}
\end{equation}
It is straightforward to demonstrate that $\C F(X)$ is non-negative. Since
$\C0(X) \ge 0$, the condition $\hat G(x,y) \le 0$ (illustrated in
Fig.~\ref{fig:homog_vs_monte_carlo}) is sufficient for the integral over the
product in \eqref{eqn:CF_solution} to be non-negative. In \eqref{eqn:G-}, the
exponential is always positive, and each hyperbolic function in $G_-(x,y)$ is
non-negative for $0 \le x, y \le \epm$ except for $\sinh(\phi(y-\epm)) \le 0$.
Therefore $G_-(x,y) \le 0$. Also, the relation \eqref{eqn:G+} only involves
swapping $x$ and $y$ and an exponential factor, so $G_+(x,y) \le 0$. Hence $\C
F(X) \ge 0$.

The correction is illustrated using the example in Fig.~\ref{fig:M=100_sims}.
We use $\ep=0.1$, implying that only limited accuracy can be expected of the
homogenization approximation, and $\Pe=O(1)$ implying that the staircase
structure appears at higher order in $\ep$. In this case $C^{(0)}$ captures the
decay in the mean concentration with distance reasonably well, while $C^{(F)}$
captures the correct sense and magnitude of the correction due to fluctuations
in sink strength.

\begin{figure*}
    \centering
    \includegraphics[width=\linewidth]{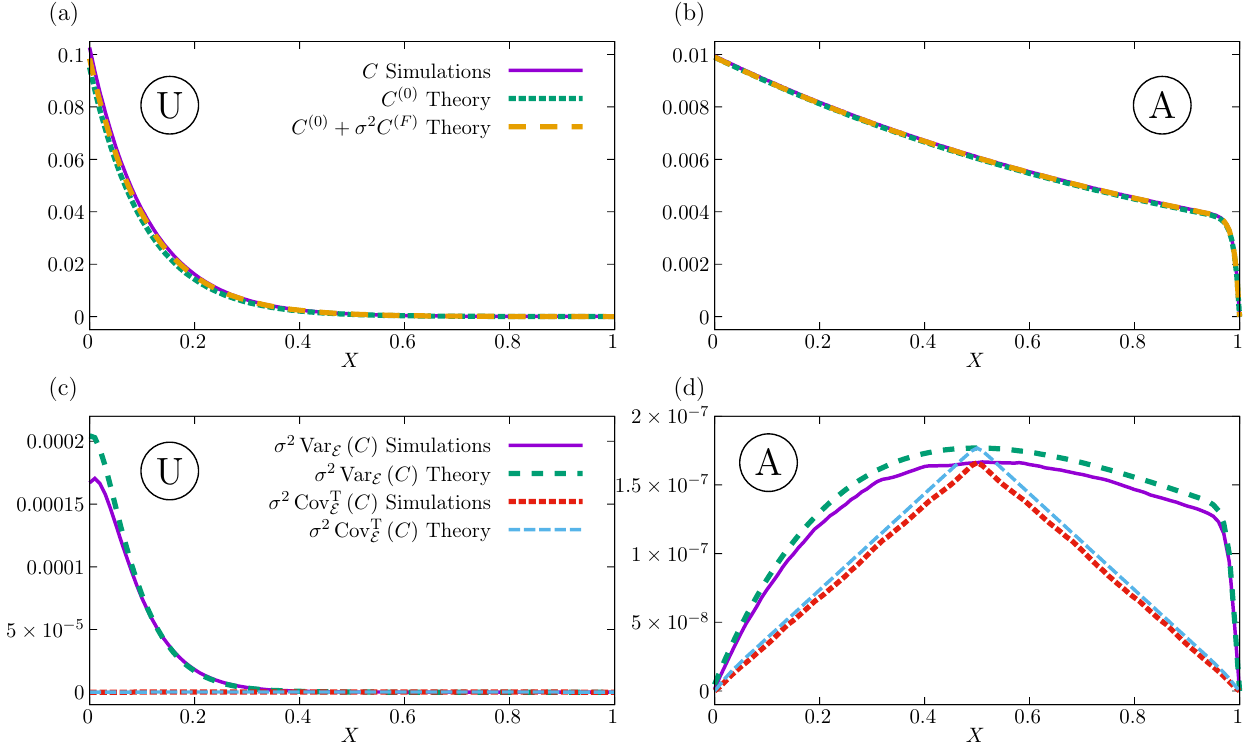}
    \caption{\label{fig:lognormal_sinks} Comparison between (a),(b) sample means
        $\sE{C(X)}$ (solid) and the theoretical prediction
        \eqref{eqn:C0_solution} (thin-dashed), (c),(d) variances $\sigma^2
        \svarnb{(\hat C(X))}$ (solid) and transverse covariances $\sigma^2
        \stcovnb{(\hat C(X))}$ (thin-dashed) and the theoretical predictions
        (\ref{eqn:C_distribution}, \ref{eqn:cov_theory}) (wide-dashed,
        medium-dashed), with sink strengths $S_i$ distributed lognormally with
        variance $\sigma^2 = 1$. In addition, the wide-dashed line in (a),(b)
        shows the prediction of the mean including the correction $\C F$ due to
        extrinsic fluctuations. Parameter values $(\Pe,\Da)$ are (a),(c)
        $(\ep,\ep)$, (b),(d) $(1,\ep)$ and all other parameters as in
        Figures~\ref{fig:homog_vs_monte_carlo},\ref{fig:homog_vs_monte_carloB}.}
\end{figure*}

Finally, to test how well this approach works for larger sink variances, we
present simulations with log-normally distributed sink strengths, ensuring that
$S_i > 0$. Figure~\ref{fig:lognormal_sinks} compares simulations with $\sigma^2
= 1$ against the theoretical predictions of the mean \eqref{eqn:C0_solution},
its correction (\ref{eqn:CF_solution}) and the covariance
\eqref{eqn:cov_theory}. The small-$\sigma$ predictions of mean and variance
provide surprisingly good approximations of both quantities. We now seek to
understand in more detail the range of validity of the approximation.

\subsection{Size of fluctuations}
\label{sec:size}

\begin{figure}
    \centering
    \includegraphics[width=0.95\columnwidth]{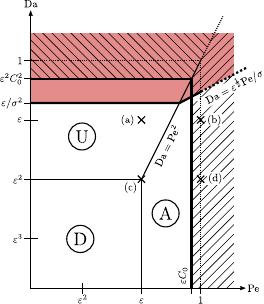}
    \caption{\label{fig:parameters}A schematic diagram of $(\Pe,\Da)$-parameter
        space, identifying asymptotic regions for which diffusion (D), uptake
        (U) and advection (A) are dominant across the whole domain (with
        diffusion being dominant between sinks). The shaded (red) region
        denotes, for illustrative values of $\ep$, $\sigma$ and $C_0$ (with $1
        \ll C_0 \ll \epm$, $\sigma^2 \ll 1 \ll \ep^{1/2}\sigma C_0$), parameter
        values for which extrinsic fluctuations are dominant, and the
        homogenization approximation does not apply. Hatching represents the
        regions in which the intrinsic noise becomes as large as the mean
        concentration. Points marked (a)--(d) correspond to the panels in Figs.
        5 and 6.}
\end{figure}

\begin{table*}
    \centering
    \begin{tabular}{ c || c | c | c | c }
        Regime & $\C0$ & $\hat G$ & $\overline{\rho_{i}^\text{st}}/C_0$ &
        $\scov{\rho_i^\text{st}}{\rho_j^\text{st}}$\\\hline\hline
        $\mathrm{[D]}$ &
        $1$ &
        $\epm$ &
        $(1 + \Da^2 \ep^{-3} \sigma^2)/C_0$ &
        $\Da^2 \ep^{-3} \sigma^2$\\\hline
        $\mathrm{[U]}$ &
        $\ep \Da^{-\frac{1}{2}}$ &
        $\Da^{-\frac{1}{2}}$ &
        $\ep \Da^{-\frac{1}{2}}\left(1 + \Da \sigma^2 \epm \right)/C_0$ &
        $\ep\sigma^2$\\\hline
        $\mathrm{[A]}$ &
        $\ep \Pe^{-1}$ &
        $\Pe^{-1}$ &
        $\ep \Pe^{-1}\left(1 + \Da^2 \sigma^2 \epm \Pe^{-2} \right)/C_0$ &
        $\Da^2 \Pe^{-4}\ep\sigma^2$
    \end{tabular}
    \caption{\label{tab:magnitudes}Estimates of magnitudes of the mean
    concentration, Green's function, total variance and extrinsic covariance in
    terms of their scaling dependence on dimensionless parameters.}
\end{table*}

It is instructive to consider the outcome of the model in various regions of the
space spanned by the parameters $\Pe$ and $\Da$. Figure~\ref{fig:parameters}
illustrates three distinct asymptotic regimes for which diffusion is dominant
between sinks. These are evident from balancing the three terms in
\eqref{eqn:c0_eqn}: (i) diffusion dominates advection and uptake for $\Pe \ll
\ep$, $\Da \ll \ep^2$; (ii) advection is dominant for $\ep\ll \Pe \ll 1$, $\Da
\ll \Pe^2$, which is the case in Figs~\ref{fig:homog_vs_monte_carlo}(b,d) and
\ref{fig:homog_vs_monte_carloB}(b,d); and (iii) uptake is dominant for
$\max(\ep^2, \Pe^2) \ll \Da \ll 1$, as in Figs~\ref{fig:homog_vs_monte_carlo}(a)
and \ref{fig:homog_vs_monte_carloB}(a). We label these regimes by circled
letters D, A and U respectively in the figures. All three effects are in balance
for $\Pe=O(\ep)$, $\Da=O(\ep^2)$; this is the case in
Figs.~\ref{fig:homog_vs_monte_carlo}(c) and \ref{fig:homog_vs_monte_carloB}(c).
For $\Pe=O(1)$ or larger, advection becomes dominant at the intersink distance;
for $\Da=O(1)$ or larger, there is complete uptake across a small number of
sinks.

We can analyse the magnitudes of the contributions to the total covariance
\eqref{eqn:total_covariance} from the intrinsic and extrinsic noise for each
parameter regime. To do so, we estimate the magnitudes of $\C0$ and $\hat G$ by
considering the dominant terms in governing equations \eqref{eqn:adre_nondim}
and \eqref{eqn:greens_func_equation} in the different regimes, and then use the
estimates $\sigma^2 C^{(F)}\sim \Da^2 \sigma^2 C^{(0)} G_{\pm}^2/\ep$ (from
\eqref{eqn:CF_solution}) and $\sigma^2 \scovnb \sim \Da^2 \sigma^2 [\C0]^2
G_{\pm}^2/\ep$ (from \eqref{eqn:cov_theory}). The homogenization approximation
fails when $\sigma^2 \C F$ becomes as large as $\C0$, or equivalently when the
extrinsic fluctuations (measured by the size of their standard deviation) become
as large as the mean concentration. We note also that $\scovnb$ should be
multiplied by $C_0^2$ and $\overline{\rho_{i}^{st}}$ by $C_0$ to transform back
to dimensionful variables; see \eqref{eq:Cscale}. As we are only interested in
the relative magnitude of mean and (co)variance we simply divide the mean by
$C_0$ in Table~\ref{tab:magnitudes}, where we summarise our results, assuming
$\sigma$ is no greater than $O(1)$. The following picture emerges.
\begin{itemize}
    \item[{[D]}] When diffusion is dominant over uptake and advection, the
        extrinsic noise is always small because $\Da^2 \sigma^2 \ll \ep^3$. The
        correction to the total mean due to extrinsic fluctuations can be
        neglected. The variance is dominated by the intrinsic noise provided
        $\Da^2 \sigma^2 C_0 \ll \ep^3$. Fluctuations due to intrinsic noise are
        small compared to the mean occupancy (\hbox{i.e.}
        $\sqrt{\overline{\rho_i^{\mathrm{st}}}/C_0}\ll C^{(0)}$) provided
        $C_0\gg 1$.
    \item[{[U]}] When uptake dominates advection (taking place over a length
        scale $x\sim \Da^{-1/2}$), the correction to the total mean due to
        extrinsic fluctuations becomes significant for $\Da \gtrsim
        \ep/\sigma^2$, implying a breakdown in the homogenization approximation;
        the example in Fig.~\ref{fig:lognormal_sinks}(a,c) sits at this
        threshold. The intrinsic noise becomes as large as the mean (\hbox{i.e.}
        $\sqrt{\overline{\rho_i^{\mathrm{st}}}/C_0}\gtrsim C^{(0)}$) for $\Da
        \gtrsim \ep^2 C_0^2$. There are therefore two independent thresholds at
        which the system becomes strongly disordered, with the size of the
        parameter $\ep^{1/2} \sigma C_0$ relative to unity determining which one
        dominates.
    \item[{[A]}] When advection dominates, $\C0$ and $\hat G$ exhibit boundary
        layers of length $x\sim 1/\Pe$. Extrinsic fluctuations become dominant
        for $\Da \sigma \gtrsim \ep^{1/2} \Pe$ (the example in
        Fig.~\ref{fig:lognormal_sinks}(b,d) sits just below this threshold) and
        intrinsic noise becomes as large as the mean for $\Pe \gtrsim \ep C_0$.
\end{itemize}
These thresholds are illustrated in Fig.~\ref{fig:parameters}. The conditions
on $C_0$ (see (\ref{eq:Cscale})) for intrinsic noise to be small compared to the
mean can be re-expressed in terms of the parameters of the discrete model as
\begin{equation}
    \label{eq:alpha}
    \alpha\gg \max\left[(p+q)/M, [S_0 (p+q)]^{1/2}, \Delta(p-q) \right],
\end{equation}
the three conditions applying in the diffusion-, uptake- and advection-dominated
regimes respectively.

\section{Discussion}

We have investigated a model transport problem that incorporates both intrinsic
noise associated with the underlying stochastic hopping process, and extrinsic
disorder arising from variability in sink strengths. The former generates
independent fluctuations in site occupancies, represented by a diagonal
covariance matrix, typical of a ZRP. The latter generates long-range
perturbations that can be correlated across the entire domain. We examined the
case in which multiple sinks are distributed sparsely across the domain,
allowing continuum multiscale approximations to be adopted. While it is natural
to predict mean site occupancies using the ensemble-averaged sink strength
(represented by the leading-order homogenized solution $\C0$), we found this to
be a biased estimator of the true ensemble mean. This is because a locally
elevated [diminished] sink strength leads to global reduction [increase] in
concentration, including at the sink itself. This in turn leads to a net
reduction in the average local uptake rate $CS$ (represented by $\langle \hat{C}
\hat{S}_j\rangle_{\mathcal{E}}\leq 0$ in (\ref{eqn:cbar_eqn})). The homogenized
solution therefore overestimates the uptake rate when there is variability in
sink strength, and therefore underestimates the mean site occupancy.

We used stochastic homogenization to derive explicit predictions of the
fluctuations arising from weak sink disorder, and validated the predicted
covariance against simulations. The transport process has three competing
physical effects -- diffusion, advection, and uptake -- and a relatively
complicated interplay between these effects is observed. The convergence of the
homogenization approximation to the ensemble mean is parameter-dependent,
weakening with increasing mean sink strength; i.e.\ with increasing $\Da$ in
Fig.~\ref{fig:parameters}. The condition for homogenization to fail,
$\sigma^2\gtrsim \max(\ep/\Da, \ep \Pe^2/\Da^2)$, can be expressed in terms of
the parameters of the discrete model as
\begin{equation}
    \label{eq:largesigma}
    \sigma^2 \gtrsim \max \left( \frac{(p+q)N}{S_0 M^2}, \frac{(p-q)^2 N}{S_0^2
    M^2} \right),
\end{equation}
which shows how the effects of disorder become important when the number of
sinks falls and their strength increases. We estimated the relative magnitudes
of intrinsic and extrinsic noise, showing how the former becomes prevalent as
the inlet flux $\alpha$ diminishes (see (\ref{eq:alpha})). Our analysis
indicates that the parameter $\ep^{1/2} \sigma C_0\sim \ep^{1/2}
\sigma \alpha M/(p+q)$ must be small compared to unity for intrinsic noise to
dominate extrinsic noise.

In the present study we have not sought to describe the case of
strong quenched disorder, defined by (\ref{eq:largesigma}) and indicated by the
shaded region in Fig.~\ref{fig:parameters}. We anticipate that individual
realisations will deviate significantly from the ensemble average, making the
system non-self-averaging in this parameter regime. Techniques from
condensed-matter physics, such as the coherent medium approach and related
methods \cite{LaxOdagaki,Bernasconi}, could be useful for estimating mean
transport properties. Likewise we have not addressed time-dependent variations
in detail, for which anomalous transport effects can be anticipated; this has
been illustrated for a related chemical transport problem in the weak disorder
regime \cite{bolsterdentz}, and framed as a continuous-time random walk
\cite{Berkowitz}.

Returning to one of our motivating problems, for oxygen transport in
a placental subunit (a placentone), the P\'eclet number has been estimated to be
of order $10^3$ to $10^4$ \cite{chern10}. Taking the domain length $L$ to be
comparable to the path length ($O(1\si{cm})$) from a spiral artery to a
draining decidual vein, this implies $\Pe/\ep\gtrsim 10^3$, a regime in
which advection dominates at the microscale. The spatial disorder of villous
branches within the placentone will contribute to fluctuations in the
concentration field induced by variability in uptake strength (as modelled
here). Intrinsic noise due to small particle numbers can be expected to be
negligible; however the influence of fluctuations in the flow field induced by
the irregular geometry may be significant \cite{deAnna13} and will be addressed
elsewhere. An alternative application for which intrinsic and extrinsic noise
may be of comparable importance concerns the motion of inhaled nanoparticles
(such as viruses or drugs) through the mucus lining of a lung airway
\cite{Lai2009,Cu2009}: here predominantly diffusive transport may be mediated by
trapping of particles by large mucin molecules. While the present model
describes a limited number of features of such applications, it provides a
framework for describing the magnitude, structure and influence of
fluctuations.

The problem we have addressed has a number of obvious extensions, including
spatially correlated or more densely distributed sinks, random sink locations,
nonlinear kinetics and nonlinear hopping rates, higher spatial dimensions, etc.
These extensions can be adapted to study specific applications in natural
systems involving transport in the presence of spatial disorder. Of particular
significance in terms of predictive modelling is understanding the
nature and magnitude of the bias in the homogenization prediction. The present
approach is a weak disorder expansion (see (\ref{eq:largesigma})) that allows
the physical system to be described as a Gaussian process with slowly varying
mean and spatial covariance. While this approach has wide applicability as a
method of uncertainty quantification, alternative approaches are needed to
address the strong disorder case in which extrinsic fluctuations appear at
leading order.

\section*{Acknowledgements}

OEJ and TG acknowledge support from EPSRC grant EP/K037145/1.

\begin{appendix}
    \section{Time evolution of the moments of the stochastic model}
    \label{app:moments}
    \subsection{Master equation and first moment}
    In this Appendix we briefly summarise the derivation of the differential
    equations for the first and second moments of the stochastic hopping model.
    This is for a fixed realisation of the sinks, and describes an average over
    the intrinsic disorder only. The derivation is standard, see e.g.
    \citep{van2007stochastic}, but it is useful to include a brief summary here.

    From the master equation \eqref{eqn:master_equation_generic}, one finds
    \begin{equation}
        \label{eqn:jump_moments_generic}
        \begin{aligned}[b]
            &\diff{}{t}\rE{f(\V n) | \V S}\\
            &= \sum_\V n f(\V n) \diff{}{t}P(\V n,t | \V S)\\
            &= \sum_\V n \sum_\V m \left[f(\V m)\right. - \left. f(\V n)\right]
            W_{\V n\to\V m}(\V S) P(\V n,t | \V S).
        \end{aligned}
    \end{equation}

    If we choose $f(\V n) = n_i$, we obtain the equations governing the
    time-evolution of the mean occupancies $\rho_i(t | \V S) = \rE{n_i | \V S}$,
    \begin{equation}
        \label{eqn:1st_moments_generic}
        \diff{}{t}\rho_i(t | \V S) = \rE{a^{(1)}_i(\V n, \V S, t) | \V S}, \quad
        i=1,\dotsc, M
    \end{equation}
    where the first jump moment at $\V n$ is defined as
    \begin{equation}
        \label{eqn:1st_jump_moment_generic}
        a^{(1)}_i(\V n, \V S) = \sum_\V m (m_i -
        n_i)W_{\V n\to\V m}(\V S).
    \end{equation}
    In our model the changes each reaction produces do not depend on the present
    state of the system (the stoichiometric coefficients are constants). The
    rates $W_{\V n\to\V m}(\V S)$ only involve constant terms and terms
    involving first power of particle numbers, but no non-linear contributions.
    The jump moments are hence of the form
    \begin{equation}
        \label{eqn:1st_jump_moment_linear}
        a^{(1)}_i(\V n,\V S) = \sum_{k=1}^M A^{(1)}_{ik}(\V S) n_k +
        B^{(1)}_i(\V S),
    \end{equation}
    with suitable coefficients $A^{(1)}_{ik}(\V S)$ and $B^{(1)}_i(\V S)$. Given
    this (affine) linear form, one then has $\rE{a^{(1)}_i(\V n, \V S) | \V S} =
    a^{(1)}_i(\V \rho(t | \V S), \V S)$, and the equations for the intrinsic
    mean of the occupancies take the form,
    \begin{equation}
        \diff{}{t}\rho_i(t | \V S) = a^{(1)}_i(\V \rho(t | \V S), \V S).
        \quad i=1,\dotsc, M.
    \end{equation}

    In the main text we refer to the matrix $\M A^{(1)}$ and vector $\V B^{(1)}$
    for our model \eqref{eqn:transition_rates} by simply $\M A$ and $\V v$.
    The first jump moments are given in (\ref{eqn:A^1_model}a), with $v_i =
    \alpha \delta_{i,1}$. Therefore the time-evolution of the mean occupancies
    is governed by \eqref{eqn:mean_occupancy} upon writing the bulk and boundary
    equations out explicitly.

    \subsection{Second moment}
    \label{sec:second}

    Now turning to the covariances, we start from
    \begin{gather}
        \label{eqn:covariances_generic_diff}
        \begin{aligned}
            \diff{}{t}\sigma_{ij}(t | \V S) &=
            \diff{}{t}\left[\rE{n_i n_j | \V S} - \rE{n_i | \V S}\rE{n_j | \V
                    S}\right]\\
            &= \diff{}{t}\rE{n_i n_j | \V S} -
            \rE{n_i | \V S}\diff{}{t}\rE{n_j | \V S}\\
            &\quad - \rE{n_j | \V S}\diff{}{t}\rE{n_i | \V S},
        \end{aligned}
    \end{gather}
    for $i,j=1,\dotsc,M$. Choosing $f(\V n) = n_i n_j$ in
    \eqref{eqn:jump_moments_generic} gives
    \begin{gather}
        \label{eqn:2nd_moments_generic}
        \begin{aligned}
            \diff{}{t}\rE{n_i n_i | \V S} &= \rE{a^{(2)}_{ij}(\V n, \V S) | \V
                S} + \rE{n_i a^{(1)}_j(\V n,\V S) | \V S}\\
            &\quad + \rE{n_j a^{(1)}_i(\V n, \V S) | \V S},
        \end{aligned}
    \end{gather}
    where the second jump moments $a^{(2)}_{ij}(\V n,\V S)$ are defined as
    \begin{equation}
        \label{eqn:2nd_jump_moment}
        a^{(2)}_{ij}(\V n,\V S) = \sum_\V m (m_i - n_i)(m_j - n_j)W_{\V n\to\V
            m}(\V S).
    \end{equation}
    Using \eqref{eqn:covariances_generic_diff} with
    \eqref{eqn:1st_moments_generic} and \eqref{eqn:2nd_moments_generic}, we can
    write the time-evolution of the covariances in terms of the first and second
    jump moments:
    \begin{equation}
        \begin{aligned}
            \label{eqn:covariances_generic}
            \diff{}{t}\sigma_{ij}(t | \V S) &= \rE{a^{(2)}_{ij}(\V n,\V S) |
                \V S}\\
            &\quad + \rE{\left(n_i-\rho_i\right) a^{(1)}_j(\V n,\V S) | \V S}\\
            &\quad + \rE{\left(n_j-\rho_j\right) a^{(1)}_i(\V n,\V S) | \V S}.
        \end{aligned}
    \end{equation}
    Noting again the linearity of the reaction rates in the particle numbers and
    the fact that the stoichiometric coefficients are constant,
    \eqref{eqn:covariances_generic} simplifies to
    \eqref{eqn:covariance_occupancy} where $\M B = a_{ij}^{(2)}(\V \rho(t|\V
    S),\V S)$ and the matrix $A^{(1)}_{ik}(\V S)$ is defined in
    \eqref{eqn:1st_jump_moment_linear}.

\section{Long-range correlation of fluctuations}
\label{sec:corr}

Using the expression \eqref{eqn:hat_C} for $\hat C(x)$, we can calculate the
spatial covariance structure of the fluctuations. We introduce $y \in [0,\epm]$
and $Y = \ep y$ as the second short- and long-range variables, and define $j =
\floor{y}$. Then, using the bilinearity of the covariance,
\begin{widetext}
    \begin{equation}
        \begin{aligned}
            \scov{\hat C(x)}{\hat C(y)} &=
            \begin{aligned}[t]
                \scovnb{
                \biggl(
                    &\Da \biggl\{ \sum_{k=1}^i \hat S_k \C0(\ep k) G_+(x,k) +
                    \sum_{k=i+1}^N \hat S_k \C0(\ep k) G_-(x,k)\biggr\},\\
                    &\Da \biggl\{ \sum_{l=1}^j \hat S_l \C0(\ep l) G_+(y,l) +
                    \sum_{l=j+1}^N \hat S_l \C0(\ep l) G_-(y,l)\biggr\}
                \biggr)}
            \end{aligned}\\
            &=
            \begin{aligned}[t]
                \Da^2 \Biggl\{
                      & \sum_{k=1}^i \sum_{l=1}^j \C0(\ep k) \C0(\ep l) G_+(x,k)
                    G_+(y,l) \scov{\hat S_k}{\hat S_l}\\
                    + & \sum_{k=1}^i \sum_{l=j+1}^N \C0(\ep k) \C0(\ep l)
                    G_+(x,k) G_-(y,l) \scov{\hat S_k}{\hat S_l}\\
                    + & \sum_{k=i+1}^N \sum_{l=1}^j \C0(\ep k) \C0(\ep l)
                    G_-(x,k) G_+(y,l) \scov{\hat S_k}{\hat S_l}\\
                    + & \sum_{k=i+1}^N \sum_{l=j+1}^N \C0(\ep k) \C0(\ep l)
                    G_-(x,k) G_-(y,l) \scov{\hat S_k}{\hat S_l}
                \Biggr\}.
            \end{aligned}
        \end{aligned}
    \end{equation}
    Since
    $\scov{\hat S_k}{\hat S_l} = \delta_{k,l}$, the covariance
    simplifies to
    \begin{equation}
        \begin{aligned}
            \scov{\hat C(x)}{\hat C(y)} &=
            \begin{aligned}[t]
                \Da^2 \Biggl\{
                \ \ & \smashoperator{\sum_{k=1}^{\min(i,j)}} \C0(\ep k)^2
                    G_+(x,k) G_+(y,k)
                +     \smashoperator{\sum_{k=j+1}^{i}} \C0(\ep k)^2 G_+(x,k)
                    G_-(y,k)\\
                +   & \smashoperator{\sum_{k=i+1}^{j}} \C0(\ep k)^2 G_-(x,k)
                    G_+(y,k)
                +     \smashoperator{\sum_{k=\max(i,j)+1}^N} \C0(\ep k)^2
                    G_-(x,k) G_-(y,k)
                \Biggr\}\\
            \end{aligned}\\
            &= \Da^2 \sum_{k=1}^N \C0(\ep k)^2 \hat G(x,k) \hat G(y,k),
        \end{aligned}
    \end{equation}
    where the piecewise nature of $\hat G$ takes care of the different sums.
    Then by approximating the above sums with integrals to leading order, we
    have the following expression for the covariance in long-range coordinates:
    \begin{equation}
        \label{eqn:cov_theory}
            \scov{\hat C(X)}{\hat C(Y)} \approx \epm \Da^2 \int_0^1
            \C0(X')^2 \hat G(\epm X, \epm X') \hat G(\epm Y, \epm X') \dd X'.
    \end{equation}
    Recall that $\hat G$ varies by $O(1)$ with respect to the slow variable $X$
    when $\Pe=O(\ep)$, $\Da=O(\ep^2)$.
\end{widetext}
\end{appendix}


%

\end{document}